\documentclass[journal,12pt,onecolumn]{IEEEtran}
\ifCLASSINFOpdf
\else
\fi
\usepackage{setspace}
\ifCLASSOPTIONonecolumn \doublespace \fi
\usepackage{amsfonts}
\usepackage{amsmath}
\usepackage{graphicx}
\graphicspath{{Figures/}} \DeclareGraphicsExtensions{.eps,.ps}
\newcounter{MYtempeqncnt}
\allowdisplaybreaks[1]

\title{Normalized Volume of Hyperball in Complex Grassmann Manifold and Its Application in Large-Scale MU-MIMO Communication Systems}

\ifCLASSOPTIONconference
\author{\IEEEauthorblockN{Dengkui Zhu \authorrefmark{1}, Boyu Li \authorrefmark{1}, and Ping Liang \authorrefmark{1}\authorrefmark{2}}
\\
\IEEEauthorblockA{\authorrefmark{1}RF DSP Inc., 30 Corporate Park, Suite 210, Irvine, CA 92606, USA,\\ e-mail: zhudengkui@gmail.com, byli@rfdsp.com, pliang@rfdsp.com} \ifCLASSOPTIONonecolumn \\ \fi 
\IEEEauthorblockA{\authorrefmark{2}Department of Electrical Engineering, University of California - Riverside, \ifCLASSOPTIONonecolumn \\ \fi Riverside, CA 92521, USA, \ifCLASSOPTIONtwocolumn \\ \fi e-mail: liang@ee.ucr.edu} 
}
\else
\author{Dengkui~Zhu,~Boyu~Li,~\IEEEmembership{Member,~IEEE,}~and~Ping~Liang
\thanks{
The authors are with RF DSP Inc., 30 Corporate Park, Suite 210, Irvine, CA 92606, USA (e-mail: zhudengkui@gmail.com; byli@rfdsp.com; pliang@rfdsp.com).

P. Liang is also with University of California - Riverside, Riverside, CA 92521, USA (e-mail: liang@ee.ucr.edu). 
}}
\fi

\begin{document}
\maketitle
\begin{abstract}
This paper provides a solution to a critical issue in large-scale Multi-User Multiple-Input Multiple-Output (MU-MIMO) communication systems: how to estimate the Signal-to-Interference-plus-Noise-Ratios (SINRs) and their expectations in MU-MIMO mode at the Base Station (BS) side when only the Channel Quality Information (CQI) in Single-User MIMO (SU-MIMO) mode and non-ideal Channel State Information (CSI) are known? A solution to this problem would be very beneficial for the BS to predict the capacity of MU-MIMO and choose the proper modulation and channel coding for MU-MIMO. To that end, this paper derives a normalized volume formula of a hyperball based on the probability density function of the canonical angle between any two points in a complex Grassmann manifold, and shows that this formula provides a solution to the aforementioned issue. It enables the capability of a BS to predict the capacity loss due to non-ideal CSI, group users in MU-MIMO mode, choose the proper modulation and channel coding, and adaptively switch between SU-MIMO and MU-MIMO modes, as well as between Conjugate Beamforming (CB) and Zero-Forcing (ZF) precoding. Numerical results are provided to verify the validity and accuracy of the solution.
\end{abstract}

\begin{IEEEkeywords}
Complex Grassmann manifold, Hyperball, MU-MIMO, Non-ideal CSI, Normalized volume formula, SINR
estimation.
\end{IEEEkeywords}



%

\ifCLASSOPTIONonecolumn \newpage \fi
\section{Introduction}
%
%
%
%
Complex Grassmann manifold has found its extensive applications in Multiple-Input Multiple-Output (MIMO) wireless communication [1]-[5] in the last ten years. Many of these applications focused on solving Grassmann packing or quantization problems for better performance in codebook design [3]-[5], where the eigenspace of the wireless channel matrix is modeled as a point in complex Grassmann manifold. There are also published works on theoretic analyses of the quantization bound on the complex Grassmann manifold [6], [7]. For the latter, the normalized volume of hyperball needs to be calculated. Various of approximated volume formulas were obtained in [6]-[11] and used extensively, where the estimation results in [8] were extensively applied and accurate enough for these applications. However, they are not suitable for some other special applications, e.g., when the normalized hyperball volume is used as the probability of the distance between any two points in complex Grassmann manifold [5], [12]. Hence, a more precise normalized volume formula of hyperball in complex Grassmann manifold is needed.

Large-scale MIMO or massive MIMO system was firstly introduced 
in [13] where the Base Station (BS) is equipped with dozens to several hundreds transmit antennas. It has received enormous attention due to its ability of providing linear capacity growth without the need of increased power or bandwidth [13], [14]. This advantage is realized by employing Multi-User MIMO (MU-MMO) which simultaneously beam-forms to many users. In this system, the BS selects users at each scheduling slot and transmits data to these users on the same time and frequency resource. In order to remove the mutual interference among these users and maximize the multi-user channel capacity, the BS needs to know the Channel State Information (CSI) and Channel Quality Information (CQI) of each user. However, in practical systems, e.g., 3rd Generation Partnership Project (3GPP) Long Term Evolution-Advanced (LTE-A) [15], there always exists unavoidable errors in CSI estimation because of the limited feedback bandwidth in Frequency-Division Duplexing (FDD) systems or the noise and interference in the measured results of Time-Division Duplexing (TDD) systems. On the other hand, the BS generally only knows the CQI of each user in Single-User MIMO (SU-MIMO) mode but not in MU-MIMO mode because it is impractical for the system to actually measure the CQI in MU-MIMO mode. Therefore, even if Zero-Forcing (ZF) precoding is used at the BS side for MU-MIMO, there is still residual interference due to inaccurate CSI in this situation. As a result, it affects MU-MIMO systems in two aspects. Firstly, the real SINR at each receiver side is greatly changed compared to the SINR of SU-MIMO mode, hence a reasonable estimation of each user's real SINR is needed so that it could be used to properly select each user's data transmission rate (e.g., the modulation scheme and channel coding rate) in MU-MIMO mode. Secondly, as the channel capacity of each user in MU-MIMO mode is smaller than SU-MIMO mode because of not only the decrease of transmit power but also the residual interference, it requires us to forecast the system throughput gain achieved by MU-MIMO cautiously. Especially, when the error of CSI is large, MU-MIMO may have no advantage compared to SU-MIMO but significantly increase system complexity. Therefore, how to choose the proper modulation order and channel coding rate and to predict the capacity of MU-MIMO with non-ideal CSI is a fundamental question in actual MU-MIMO communication systems. However, there are only a few papers analyzing the real SINR and capacity for MU-MIMO in practical systems. In [16], a cursory SINR prediction scheme was presented for some specific scenarios. In [17], the Cumulative Density Function (CDF) of SINR with ideal CSI for MU-MIMO was derived but the errors in practical systems were not considered.

In this paper, precise normalized volume formulas of hyperball based on two general distance definitions (projective-Frobenius norm and projective-2 norm) are obtained by applying the probability density function of canonical angle between any two points in complex Grassmann manifold introduced in [18]. Simulation results show that they have very high accuracy thus verified their reasonableness. One of the formulas is applied to estimate the SINRs in large-scale MU-MIMO communication systems based on SU-MIMO SINRs. In various simulated cases, our analytical expressions of SINRs match the real values in the actual system almost perfectly. It indicates that these expressions could be used to forecast the MU-MIMO SINRs and the capacity gain over SU-MIMO.

The rest of this paper is organized as follows. Firstly, the foundation of complex Grassmann manifold is introduced in Section I. The hyperball volume formula for various cases based on different definitions of norm distances is proposed in Section II, and simulation results are given to demonstrate accuracy. In Section III, we apply the formula to estimate the SINRs in large-scale MU-MIMO systems and approximation expressions are derived. Simulation results are provided for verification. Finally, in Section IV, conclusions are drawn.

\section{Foundation Of Complex Grassmann Manifold}
In this section, we first provide the definitions of complex Grassmann manifold and the distance on it. Then, the normalized volume of hyperball in Grassmann manifold is introduced, and the equivalent relation between CDF of distance in the manifold and normalized volume of hyperball is constructed. Finally, the normalized volume formula is obtained for various cases.

\subsection	{Complex Grassmann Manifold}
In complex vector space $\mathbb{C}^n $, the set of all $n \times k $ matrices with mutually orthogonal columns where each column has unit norm are defined as complex Stiefel manifold, i.e.,
\begin{equation}
\label{f1}
{\mathbb S}{\mathbb T}\left( {k,n} \right) = \left\{ {\left. {{\bf{V}} \in {\mathbb C}^{n \times k} } \right|{\bf{V}}^H {\bf{V}} = {\bf{I}}_k } \right\}.
\end{equation}
An equivalent relation between ${\bf{V}}_1 $ and ${\bf{V}}_2 $ is constructed on ${\mathbb S}{\mathbb T}\left( {k,n} \right)$ as $ {{\bf{V}}_1} \sim {{\bf{V}}_2} $ for any $ {{\bf{V}}_1},{{\bf{V}}_2} \in {\mathbb S}{\mathbb T}\left( {k,n} \right) $ if ${\bf{V}}_1  = {\bf{V}}_2 {\bf{U}}$, where ${\bf{U}} \in {\mathbb U}\left( k \right)$ and ${\mathbb U}\left( k \right)$ is the unitary group with order $k$. Based on the equivalent relation, a quotient space of ${\mathbb S}{\mathbb T}\left( {k,n} \right)$ is expressed as
\begin{equation}
\label{f2}
{\mathbb G}\left( {k,n} \right) = {{{\mathbb S}{\mathbb T}\left( {k,n} \right)} \mathord{\left/
 {\vphantom {{{\mathbb S}{\mathbb T}\left( {k,n} \right)} {{\mathbb U}\left( k \right)}}} \right.
 \kern-\nulldelimiterspace} {{\mathbb U}\left( k \right)}}.
\end{equation}
We call the quotient space $ {\mathbb G}\left( {k,n} \right) $ complex Grassmann manifold. In this paper, a point is used to represent a matrix in ${\mathbb S}{\mathbb T}\left( {k,n} \right)$ or ${\mathbb G}\left( {k,n} \right)$, then all the points which are equivalent in ${\mathbb S}{\mathbb T}\left( {k,n} \right)$ are denoted by anyone of them. If we call this point the effective point, then $ {\mathbb G}\left( {k,n} \right) $ is made of all the effective points in ${\mathbb S}{\mathbb T}\left( {k,n} \right)$. According to the definition, a point in $ {\mathbb G}\left( {k,n} \right) $ is a $ k $-dimensional subspace of $ {{\mathbb C}^n} $, while all the equivalent points in ${\mathbb S}{\mathbb T}\left( {k,n} \right)$  represent different bases for the same $ k $-dimensional subspace.

\subsection{Distance on $ {\mathbb G}\left( {k,n} \right) $}
In order to study the volume on $ {\mathbb G}\left( {k,n} \right) $, we have to define the distance on it. Other than the general properties of distance, e.g., non-negativity and triangle inequality, an additional property that the distance between any two equivalent points is zero has to be satisfied for the defined distance. Next, we provide two common definitions.

\emph {Definition 1:} (Projective-Frobenius norm distance) For any two points $ {{\bf{V}}_1},{{\bf{V}}_2} \in {\mathbb G}\left( {k,n} \right)$, the projective-Frobenius norm distance between them is
\begin{equation}
\label{f3}
{d_{pF}} = \frac{1}{{\sqrt 2 }}{\left\| {{{\bf{V}}_1}{\bf{V}}_1^H - {{\bf{V}}_2}{\bf{V}}_2^H} \right\|_F} = \sqrt {\sum\limits_{i = 1}^k {{{\sin }^2}{\theta _i}} }
\end{equation}
where $ 0 \leq {\theta _1}, \cdots ,{\theta _k} \leq {\pi  \mathord{\left/
 {\vphantom {\pi  2}} \right.
 \kern-\nulldelimiterspace} 2} $ are the canonical angles of ${{\bf{V}}_1}  $ and $ {{\bf{V}}_2} $, which are defined by the Singular Value Decomposition (SVD) of ${\bf{V}}_1^H{{\bf{V}}_2} $
\begin{equation}
\label{f4}
{\bf{V}}_1^H {\bf{V}}_2  = {\bf{U}}_L \mathbf{\Sigma} {\bf{U}}_R 
\end{equation}
where ${\mathbf{\Sigma}}  = {\text{diag}}\left( {\cos \theta _1 , \cdots ,\cos \theta _k } \right) = \cos {\mathbf{\Theta}} $, and $ {\theta _i} = {\Theta _{i,i}},i = 1, \cdots ,k $. Obviously, $ {d_{pF}} $  satisfy the inequality $ 0 \leq {d_{pF}} \leq \sqrt k  $.

\emph {Definition 2:} (Projective-2 norm distance) For any two points $ {{\bf{V}}_1},{{\bf{V}}_2} \in {\mathbb G}\left( {k,n} \right) $, the projective-2 norm distance between them is
\begin{equation}
\label{f5}
{d_{p2}} = {\left\| {{{\bf{V}}_1}{\bf{V}}_1^H - {{\bf{V}}_2}{\bf{V}}_2^H} \right\|_2} = \max \left( {\sin {\theta _i}} \right),i = 1, \cdots ,k
\end{equation}
where $ {\theta _i},i = 1, \cdots ,k $, are the same as in Definition 1. Obviously $ {d_{pF}} $ satisfies inequality $ 0 \leq {d_{p2}} \leq 1 $.

Based on the above definitions, if we know the canonical angles of any two points in $ {\mathbb G}\left( {k,n} \right) $, it is easy to calculate the distances between them.

\subsection{Normalized Volume of Hyperball in ${\mathbb G}\left( {k,n} \right)$} 
In $ {\mathbb G}\left( {k,n} \right) $, the hyperball $ {\mathfrak{B}_{k,n}}\left( \delta  \right) $ with center $ {\mathbf{C}} $ and radius $ \delta  $ is defined as
\begin{equation}
\label{f6}
{\mathfrak{B}_{k,n}}\left( \delta  \right): = \left\{ {\left. {d\left( {{\mathbf{X,C}}} \right) \leq \delta } \right|{\mathbf{X}} \in \mathbb{G}\left( {k,n} \right)} \right\}
\end{equation}
where $ d\left( {{\mathbf{X,C}}} \right) $ is one of the two defined distances in (\ref{f4}) and (\ref{f5}). Let the volumes of $ {\mathfrak{B}_{k,n}}\left( \delta  \right) $ and ${\mathbb G}\left( {k,n} \right)$ be $ {V_d}\left( {{\mathfrak{B}_{k,n}}\left( \delta  \right)} \right) $ and $
 V_\mathbb{C} \left( {\mathbb{G}_{k,n} } \right)$ respectively, then we define the normalized volume of hyperball ${\mathfrak{B}_{k,n}}\left( \delta  \right) $ as 
\begin{equation}
\label{f7}
\bar V_d \left( {\mathfrak{B}_{k,n} \left( \delta  \right)} \right) = \frac{{V_d\left( {\mathfrak{B}_{k.n} \left( \delta  \right)} \right)}}
{{V_\mathbb{C} \left( {G_{k,n} } \right)}}.
\end{equation}
According to the Haar measure theory of compact manifold [19], this normalized volume defines the probability that the distance between any point and the center $ {\mathbf{C}} $ is no more than $ \delta $, i.e.,

\begin{equation}
\label{f8}
p\left( {d \leq \delta } \right) = \frac{{V_d\left( {\mathfrak{B}_{k,n} \left( \delta  \right)} \right)}}
{{V_\mathbb{C} \left( {G_{k,n} } \right)}} = \bar V_d \left( {\mathfrak{B}_{k,n} \left( \delta  \right)} \right).
\end{equation}
Because of the unique right unitary transformation invariant measure [19] on ${\mathbb G}\left( {k,n} \right)$, Equation (\ref{f8}) implies that $ \bar V_d \left( {\mathfrak{B}_{k,n} \left( \delta  \right)} \right) $ is exactly the probability of the distance between any two points which is no more than $ \delta  $. As a result, we could calculate the normalized volume of hyperball by the probability $ p\left( {d \leq \delta } \right) $.

\section{Normalized Volume on $ {\mathbb G}\left( {k,n} \right) $}
When $ n \geq 2k\ $, the probability density function of the canonical angles between any two points in $ {\mathbb G}\left( {k,n} \right) $ are provided in Reference [18] as
\ifCLASSOPTIONonecolumn
\begin{equation}
\label{f9}
{p_\Theta }\left( {{{\sin }^2}{\theta _1}, \cdots ,{{\sin }^2}{\theta _k}} \right) = \frac{{{\pi ^{k\left( {k - 1} \right)}}C{\Gamma _k}\left( n \right)}}{{C\Gamma _k^2\left( k \right)C{\Gamma _k}\left( {n - k} \right)}}{\left( {\prod\limits_{i = 1}^k {{{\sin }^2}{\theta _i}} } \right)^{n - 2k}}\prod\limits_{i < j}^k {{{\left( {{{\sin }^2}{\theta _i} - {{\sin }^2}{\theta _j}} \right)}^2}} 
\end{equation}
\else
\begin{align}
\label{f9}
{p_\Theta }\left( {{{\sin }^2}{\theta _1}, \cdots ,{{\sin }^2}{\theta _k}} \right) = \frac{{{\pi ^{k\left( {k - 1} \right)}}C{\Gamma _k}\left( n \right)}}{{C\Gamma _k^2\left( k \right)C{\Gamma _k}\left( {n - k} \right)}} & \nonumber \\ \times {\left( {\prod\limits_{i = 1}^k {{{\sin }^2}{\theta _i}} } \right)^{n - 2k}} \prod\limits_{i < j}^k {{{\left( {{{\sin }^2}{\theta _i} - {{\sin }^2}{\theta _j}} \right)}^2}} 
\end{align}
\fi
where $ C{\Gamma _k}\left( n \right) = {\pi ^{{{k\left( {k - 1} \right)} \mathord{\left/ {\vphantom {{k\left( {k - 1} \right)} 2}} \right. \kern-\nulldelimiterspace} 2}}}\prod\limits_{i = 1}^k {{\Gamma }\left( {n - i + 1} \right)} $ and $ \Gamma \left( n \right) = \left( {n - 1} \right)!$. Notice that $ {\sin ^2}{\theta _1}, \cdots , {\sin ^2}{\theta _k} $ are ordered with $ {\sin ^2}{\theta _1} \geq  \cdots  \geq {\sin ^2}{\theta _k} $. According to (4), two sets of different ordered angles correspond to two equivalent points in $ {\mathbb G}\left( {k,n} \right) $. With formula (\ref{f9}), combining the distances defined in (\ref{f3}) and (\ref{f5}), we write the volume of $ {\mathfrak{B}_{k,n}}\left( \delta  \right) $ in $ \mathbb{G}\left( {k,n} \right) $ as
\ifCLASSOPTIONonecolumn
\begin{equation}
\label{f10}
{\bar V_d}\left( {{\mathfrak{B}_{k,n}}\left( \delta  \right)} \right) = p\left( {d \leq \delta } \right) = M\int_{\mathbf{\Omega }} {\prod\limits_{i = 1}^k {{{\left( {{{\sin }^2}{\theta _i}} \right)}^{n - 2k}}} \prod\limits_{1 \leq i < j}^k {{{\left( {{{\sin }^2}{\theta _i} - {{\sin }^2}{\theta _j}} \right)}^2}} d{{\sin }^2}{\theta _1} \cdots d{{\sin }^2}{\theta _k}}
\end{equation}
\else
\begin{align}
\label{f10}
{\bar V_d}\left( {{\mathfrak{B}_{k,n}}\left( \delta  \right)} \right) & = p\left( {d \leq \delta } \right) \nonumber \\ 
& = M\int_{\mathbf{\Omega }} \prod\limits_{i = 1}^k {{{\left( {{{\sin }^2}{\theta _i}} \right)}^{n - 2k}}} \nonumber \\ 
& \, \, \, \times \prod\limits_{1 \leq i < j}^k {{{\left( {{{\sin }^2}{\theta _i} - {{\sin }^2}{\theta _j}} \right)}^2}} d{{\sin }^2}{\theta _1} \cdots d{{\sin }^2}{\theta _k}
\end{align}
\fi
where $ M = \frac{{{\pi ^{k\left( {k - 1} \right)}}C{\Gamma _k}\left( n \right)}}{{C\Gamma _k^2\left( k \right)C{\Gamma _k}\left( {n - k} \right)}}$. For the projective-F norm distance, the domain of integration is 
\ifCLASSOPTIONonecolumn
\begin{equation}
\label{f11}
{{\mathbf{\Omega }}_{pF}} = \left\{ {\left. {\left( {{{\sin }^2}{\theta _1}, \cdots ,{{\sin }^2}{\theta _k}} \right)} \right|\sum\limits_{i = 1}^k {{{\sin }^2}{\theta _i}}  \leq {\delta ^2},{{\sin }^2}{\theta _k} \leq  \cdots  \leq {{\sin }^2}{\theta _1} \leq 1} \right\}.
\end{equation}
\else
\begin{align}
\label{f11}
{{\mathbf{\Omega }}_{pF}} = \{ \left. {\left( {{{\sin }^2}{\theta _1}, \cdots ,{{\sin }^2}{\theta _k}} \right)} \right| \sum\limits_{i = 1}^k {{{\sin }^2}{\theta _i}}  \leq {\delta ^2}, \nonumber \\ {{\sin }^2}{\theta _k} \leq  \cdots  \leq {{\sin }^2}{\theta _1} \leq 1 \}.
\end{align}
\fi
For the projective-2 norm distance, the domain of integration is 
\ifCLASSOPTIONonecolumn
\begin{equation}
\label{f12}
{{\mathbf{\Omega }}_{p2}} = \left\{ {\left. {\left( {{{\sin }^2}{\theta _1}, \cdots ,{{\sin }^2}{\theta _k}} \right)} \right|\sin {\theta _1} \leq \delta ,0 \leq {{\sin }^2}{\theta _k} \leq  \cdots  \leq {{\sin }^2}{\theta _1}} \right\}.
\end{equation}
\else
\begin{align}
\label{f12}
{{\mathbf{\Omega }}_{p2}} = \{ \left. \left( {{\sin }^2}{\theta _1}, \cdots ,{{\sin }^2}{\theta _k} \right) \right|\sin {\theta _1} \leq \delta , \nonumber \\ 0 \leq {{\sin }^2}{\theta _k} \leq  \cdots  \leq {{\sin }^2}{\theta _1} \}.
\end{align}
\fi
Next, we provide the results for various cases of $ k $ and $ \delta $ based on the two distances.

\subsection{Case of $ k = 1 $} 
When $ k = 1 $, a point in $ \mathbb{G}\left( {1,n} \right) $ is a complex vector with unit norm in $ {\mathbb{C}^n} $, then the canonical angle $ \theta $ between two points is the usual Euclidean angle of two vectors. Both of the projective-F norm distance and projective-2 norm distance degenerate to the same as $ {d_{pF}} = {d_{p2}} = d = \sin {\theta _1} $. Substituting it into (\ref{f9}) results in
\begin{equation}
\label{f13}
p_\theta  \left( {d_{pF}^2 } \right) = \left( {n - 1} \right)\left( {d_{pF}^2 } \right)^{n - 2}.
\end{equation}
Hence, the volume of ${\mathfrak{B}_{1,n}}\left( \delta  \right) $ in $ \mathbb{G}\left( {k,n} \right)$ is integrated as
\ifCLASSOPTIONonecolumn
\begin{equation}
\label{f14}
p\left( {d_{pF}^{} \leq {\delta ^{}}} \right) = p\left( {d_{pF}^2 \leq {\delta ^2}} \right) = \int {\left( {n - 1} \right){{\left( {d_{pF}^2} \right)}^{n - 2}}d\left( {d_{pF}^2} \right)}  = {\left( \delta  \right)^{2n - 2}}.
\end{equation}
\else
\begin{align}
\label{f14}
p\left( {d_{pF}^{} \leq {\delta ^{}}} \right) & = p\left( {d_{pF}^2 \leq {\delta ^2}} \right) \nonumber \\
& = \int {\left( {n - 1} \right){{\left( {d_{pF}^2} \right)}^{n - 2}}d\left( {d_{pF}^2} \right)}  \nonumber \\
& = {\left( \delta  \right)^{2n - 2}}.
\end{align}
\fi
Moreover, as the correlation coefficient of two unit-norm vectors in ${\mathbb{C}^n}$ is $ r\left(n\right) = \sqrt {1 - d_{pF}^2}  = \cos {\theta _1}$, the CDF of the correlation coefficient could be acquired from (\ref{f14}) as
\ifCLASSOPTIONonecolumn
\begin{equation}
\label{f15}
p\left( {r\left(n\right) \leq c} \right) = p\left( {d \geq \left( {\sqrt {1 - {c^2}} } \right)} \right) = 1 - p\left( {d \leq \left( {\sqrt {1 - {c^2}} } \right)} \right) = 1 - {\left( {1 - {c^2}} \right)^{n - 1}}.
\end{equation}
\else
\begin{align}
\label{f15}
p\left( {r\left(n\right) \leq c} \right) & = p\left( {d \geq \left( {\sqrt {1 - {c^2}} } \right)} \right) \nonumber \\ & = 1 - p\left( {d \leq \left( {\sqrt {1 - {c^2}} } \right)} \right) \nonumber \\ & = 1 - {\left( {1 - {c^2}} \right)^{n - 1}}.
\end{align}
\fi

\subsection{Case of $ k = 2 $} 
When $k=2$, the distances defined in (\ref{f3}) and (\ref{f5}) are not the same anymore. We next calculate the normalized volume of the hyperball respectively.

\subsubsection{The Projective-F Norm Distance}
Due to the fact that ${d_{pF}} = \sqrt {{{\sin }^2}{\theta _1} + {{\sin }^2}{\theta _2}}$, combining (\ref{f10}) and (\ref{f11}), the volume of hyperball ${{\mathfrak{B}_{2,n}}\left( \delta  \right)}$ in $ \mathbb{G}\left( {2,n} \right)$ is 
\ifCLASSOPTIONonecolumn
\begin{equation}
\label{f16}
p\left( {d_{pF}  \leq \delta } \right) = p\left( {d_{pF}^2  \leq \delta ^2 } \right) = M\iint\limits_\Omega  {\sin ^{n - 4} \theta _1 \sin ^{n - 4} \theta _2 \left( {\sin ^2 \theta _i  - \sin ^2 \theta _j } \right)^2 d\sin ^2 \theta _1 d\sin ^2 \theta _2 }
\end{equation}
\else
\begin{align}
\label{f16}
p\left( {d_{pF}  \leq \delta } \right) & = p\left( {d_{pF}^2  \leq \delta ^2 } \right) \nonumber \\ & = M\iint\limits_\Omega  \sin ^{n - 4} \theta _1 \sin ^{n - 4} \theta _2 \nonumber \\ & \, \, \, \times \left( {\sin ^2 \theta _i  - \sin ^2 \theta _j } \right)^2 d\sin ^2 \theta _1 d\sin ^2 \theta _2 
\end{align}
\fi
where the integration is taken over
\ifCLASSOPTIONonecolumn
\begin{equation}
\label{f17}
{\mathbf{\Omega }} = \left\{ {{{\sin }^2}{\theta _1} + {{\sin }^2}{\theta _2} \leq {\delta ^2},{{\sin }^2}{\theta _2} \leq {{\sin }^2}{\theta _1},0 \leq {{\sin }^2}{\theta _1},{{\sin }^2}{\theta _2} \leq 1} \right\}.
\end{equation}
\else
\begin{align}
\label{f17}
{\mathbf{\Omega }} = \{ {{\sin }^2}{\theta _1} + {{\sin }^2}{\theta _2} \leq {\delta ^2},{{\sin }^2}{\theta _2} \leq {{\sin }^2}{\theta _1}, \nonumber \\ 0 \leq {{\sin }^2}{\theta _1},{{\sin }^2}{\theta _2} \leq 1 \}.
\end{align}
\fi
Since this is a complicated problem, and the result depends on the value of $\delta$, we proide the result as a theorem below. The details are presented in Appendix A.
\newtheorem{theorem}{Theorem}
\begin{theorem}
The normalized volume of hyperball ${\mathfrak{B}_{2,n}}\left( \delta  \right)$ in complex Grassmann manifold $\mathbb{G}\left( {2,n} \right)$ is
\end{theorem}
\begin{equation}
\label{f18}
p\left(d_{pF} \leq \delta\right) = \left\{\begin{array}{cc} \left( {\frac{\delta^2}
2} \right)^{2n - 4}  + MP_1 \left( n \right), & \delta \leq 1 \\ \left( {\frac{\delta ^2}
2} \right)^{2n - 4}  + MP_2 \left( n \right), & 1 \leq \delta \leq \sqrt 2  \end{array} \right.
\end{equation}
where 
\ifCLASSOPTIONonecolumn
\begin{equation}
\label{f19}
P_1 \left( n \right) = \delta ^{4n - 8} \left( {\frac{1}
{{n - 3}}\bar B\left( {\frac{1}
{2},n - 1,n - 2} \right) + \frac{1}
{{n - 1}}\bar B\left( {\frac{1}
{2},n - 3,n} \right) - \frac{2}
{{n - 2}}\bar B\left( {\frac{1}
{2},n - 1,n - 2} \right)} \right)
\end{equation}
\else
\begin{align}
\label{f19}
P_1 \left( n \right) & = \delta ^{4n - 8} \left( \frac{1}
{{n - 3}}\bar B\left( {\frac{1}
{2},n - 1,n - 2} \right) \right. \nonumber \\ & \,\,\, + \frac{1}
{{n - 1}}\bar B\left( {\frac{1}
{2},n - 3,n} \right) \nonumber \\ & \left. \,\,\,- \frac{2}
{{n - 2}}\bar B\left( {\frac{1}
{2},n - 1,n - 2} \right) \right)
\end{align}
\fi
and
\ifCLASSOPTIONonecolumn
\begin{align}
\label{f20}
P_2 \left( n \right) = & \delta^{4n - 8} \left( \frac{1}{n - 3} \bar B_1 \left( \frac{1}
{{\delta ^2 }},n - 1,n - 2 \right) + \frac{1}
{{n - 1}}\bar B_1 \left( \frac{1}
{{\delta ^2 }},n - 3,n \right) \right. \nonumber \\ & \left. - \frac{2}
{{n - 2}}\bar B_1 \left( \frac{1}
{{\delta ^2 }},n - 1,n - 2 \right) \right).
\end{align}
\else
\begin{align}
\label{f20}
P_2 \left( n \right) = & \delta^{4n - 8} \left( \frac{1}{n - 3} \bar B_1 \left( \frac{1}{\delta ^2 },n - 1,n - 2 \right) \right. \nonumber \\ 
& \,\,\,+ \frac{1}{n - 1}\bar B_1 \left( \frac{1}{\delta ^2 },n - 3,n \right) \nonumber \\ 
& \,\,\, \left. - \frac{2}{n - 2}\bar B_1 \left( \frac{1}{\delta ^2 },n - 1,n - 2 \right) \right).
\end{align}
\fi
In (\ref{f19}), $\bar B\left( {\alpha ,m,n} \right) = B\left( {m,n} \right) - B\left( {\alpha ,m,n} \right)$ is the beta difference function, where $B\left( {m,n} \right)$ is the beta function defined as
\begin{equation}
\label{f21}
B\left( {m,n} \right) = \int_0^1 {x^{m - 1} \left( {1 - x} \right)^{n - 1} } dx = \frac{{\left( {m - 1} \right)!\left( {n - 1} \right)!}}{{\left( {m + n - 1} \right)!}}
\end{equation}
 and $ B\left( {\alpha ,m,n} \right) = \int_0^\alpha  {{x^{m - 1}}{{\left( {1 - x} \right)}^{n - 1}}} dx$ denotes the incomplete beta function, which relates to the beta function as
\begin{equation}
\label{f22}
B\left( {\alpha ,m,n} \right) = {I_\alpha }\left( {m,n} \right)B\left( {m,n} \right)
\end{equation}
where 
\begin{equation}
\label{f23}
{I_\alpha }\left( {m,n} \right) = \sum\limits_{j = m}^{m + n - 1} {\frac{{\left( {m + n - 1} \right)!}}{{j!\left( {m + n - 1 - j} \right)!}}{\alpha ^j}} {\left( {1 - \alpha } \right)^{m + n - 1 - j}}.
\end{equation}
In (\ref{f20}), ${\bar B_1}\left( {\alpha ,m,n} \right) = \bar B\left( {\alpha ,m,n} \right) - \bar B\left( {\frac{1}{2},m,n} \right)$ is the incomplete beta difference function. The constant $M$ in (\ref{f18}) is 
\begin{equation}
\label{f24}
M = \left( {n - 1} \right){\left( {n - 2} \right)^2}\left( {n - 3} \right).
\end{equation}

\subsubsection{The Projective-2 Norm Distance}
for the projective-2 norm distance, we have $ {d_{p2}} = \sin {\theta _1}$. Combining (\ref{f10}) and (\ref{f12}), the normalized volume of ${{\mathfrak{B}_{2,n}}\left( \delta  \right)}$ in $ \mathbb{G}\left( {2,n} \right)$ is
\ifCLASSOPTIONonecolumn
\begin{equation}
\label{f25}
p\left( {d_{p2}  \leq \delta } \right) = p\left( {d_{p2}^2  \leq \delta ^2 } \right) = M\iint\limits_\Omega  {\sin ^{2\left( {n - 4} \right)} \theta _1 \sin ^{2\left( {n - 4} \right)} \theta _2 \left( {\sin ^2 \theta _1  - \sin ^2 \theta _2 } \right)^2 d\sin ^2 \theta _1 d\sin ^2 \theta _2 }
\end{equation}
\else
\begin{align}
\label{f25}
p\left( {d_{p2}  \leq \delta } \right) & = p\left( {d_{p2}^2  \leq \delta ^2 } \right) \nonumber \\ & = M\iint\limits_\Omega  \sin ^{2\left( {n - 4} \right)} \theta _1 \sin ^{2\left( {n - 4} \right)} \theta _2 \nonumber \\ 
& \,\,\, \times \left( {\sin ^2 \theta _1  - \sin ^2 \theta _2 } \right)^2 d\sin ^2 \theta _1 d\sin ^2 \theta _2 
\end{align}
\fi
where the integration is taken over
\ifCLASSOPTIONonecolumn
\begin{equation}
\label{f26}
{\mathbf{\Omega }} = \left\{ {{{\sin }^2}{\theta _1} \leq {\delta ^2},{{\sin }^2}{\theta _2} \leq {{\sin }^2}{\theta _1},0 \leq {{\sin }^2}{\theta _1},{{\sin }^2}{\theta _2} \leq 1} \right\}.
\end{equation}
\else
\begin{align}
\label{f26}
{\mathbf{\Omega }} = \{ {{\sin }^2}{\theta _1} \leq {\delta ^2},{{\sin }^2}{\theta _2} \leq {{\sin }^2}{\theta _1},\nonumber \\ 0 \leq {{\sin }^2}{\theta _1},{{\sin }^2}{\theta _2} \leq 1 \}.
\end{align}
\fi
Substituting the variables $ {x_1} = {\sin ^2}{\theta _1}$ and ${x_2} = {\sin ^2}{\theta _2}$, after derivation, the result becomes
\begin{equation}
\label{f27}
p\left( {{d_{p2}} \leq \delta } \right) = {\delta ^{4n - 8}}.
\end{equation}
This means ${\bar V_{p2}}\left( {{\mathfrak{B}_{2,n}}\left( \delta  \right)} \right) = {\delta ^{4n - 8}}$.

\subsection{Case of $ k\geq 3 $}
When $ k\geq 3 $, the situation becomes more complicated if we use the same method as $k=2$, because it is very hard to determine the domain of integration as $\delta$ changes. Before further analysis, we introduce the Selberg beta integration and generalized Selberg beta integration [20] as two lemmas which are the keys to solve the problem.

\emph {Lemma 1:} (Selberg Beta integration [20]) Let the set $ {\mathbf{\Omega }} = \left\{ {\left. {\sum\limits_{i = 1}^n {{t_i}}  \leq 1} \right|{t_i} \geq 0, i = 1, \cdots ,n} \right\}$ be the integration domain, and the parameters $\alpha$, $\beta$, and $\gamma$ satisfy $\Re \mathfrak{e}\left( \alpha  \right) > 0$, $\Re \mathfrak{e}\left( \beta  \right) > 0$, and $ \Re \mathfrak{e}\left( \gamma  \right) >  - \min \{ 1/n, {\Re \mathfrak{e}\left( \alpha  \right)} / {\Re \mathfrak{e}\left( n-1  \right)} \}$ respectively, 
then the Selberg beta integration is
\ifCLASSOPTIONonecolumn
\begin{equation}
\label{f28}
\begin{split}
&\int_{\mathbf{\Omega }} {\prod\limits_{i = 1}^n {t_i^{\alpha  - 1}{{\left( {1 - {t_i}} \right)}^{\beta  - 1}}} \prod\limits_{1 \leq i < j}^n {{{\left( {{t_i} - {t_j}} \right)}^{2\gamma }}} d{t_1}}  \cdots d{t_n}\\ 
& = \frac{{\Gamma \left( \beta  \right)}}{{\Gamma \left( {\alpha n + \beta  + n\left( {n - 1} \right)\gamma } \right)}}\prod\limits_{j = 0}^{n - 1} {\frac{{\Gamma \left( {\alpha  + j\gamma } \right)\Gamma \left( {1 + \left( {j + 1} \right)\gamma } \right)}}{{\Gamma \left( {1 + \gamma } \right)}}}.
\end{split}
\end{equation}
\else
\begin{align}
\label{f28}
&\int_{\mathbf{\Omega }} {\prod\limits_{i = 1}^n {t_i^{\alpha  - 1}{{\left( {1 - {t_i}} \right)}^{\beta  - 1}}} \prod\limits_{1 \leq i < j}^n {{{\left( {{t_i} - {t_j}} \right)}^{2\gamma }}} d{t_1}}  \cdots d{t_n} \nonumber \\ 
& = \frac{{\Gamma \left( \beta  \right)}}{{\Gamma \left( {\alpha n + \beta  + n\left( {n - 1} \right)\gamma } \right)}} \nonumber \\ 
& \,\,\, \times \prod\limits_{j = 0}^{n - 1} {\frac{{\Gamma \left( {\alpha  + j\gamma } \right)\Gamma \left( {1 + \left( {j + 1} \right)\gamma } \right)}}{{\Gamma \left( {1 + \gamma } \right)}}}.
\end{align}
\fi
The generalization of the above integration is the generalized Selberg integration stated in Lemma 2.

\emph {Lemma 2:} (Generalized Selberg Beta integration [20]) Let ${\mathbf{\Omega }} = \left\{ {0 \leq {t_i} \leq 1,i = 1, \cdots ,k} \right\}$ be the domain of integration, and the parameters $\alpha$, $\beta$, and $\gamma $ satisfy $\Re \mathfrak{e}\left( \alpha  \right) > 0$, $\Re \mathfrak{e}\left( \alpha  \right) > 0$,
$ \Re \mathfrak{e}\left( \gamma  \right) >  - \min \left\{ {{1 \mathord{\left/
 {\vphantom {1 n}} \right.
 \kern-\nulldelimiterspace} n},{{\Re \mathfrak{e}\left( \alpha  \right)} \mathord{\left/
 {\vphantom {{\Re \mathfrak{e}\left( \alpha  \right)} {\left( {n - 1} \right)}}} \right.
 \kern-\nulldelimiterspace} {\left( {n - 1} \right)}},{{\Re \mathfrak{e}\left( \beta  \right)} \mathord{\left/
 {\vphantom {{\Re \mathfrak{e}\left( \beta  \right)} {\left( {n - 1} \right)}}} \right.
 \kern-\nulldelimiterspace} {\left( {n - 1} \right)}}} \right\}$ respectively, 
 then the generalized selberg beta integral is 
\begin{equation}
\label{f29}
\begin{split}
&\int_{\mathbf{\Omega }} {\prod\limits_{i = 1}^k {t_i^{\alpha  - 1}{{\left( {1 - \sum\limits_{i = 1}^k {{t_i}} } \right)}^{\beta  - 1}}} \prod\limits_{1 \leq i < j \leq k} {{{\left( {{t_i} - {t_j}} \right)}^{2\gamma }}} d{t_1}}  \cdots d{t_n}\\
& = \prod\limits_{j = 0}^{k - 1} {\frac{{\Gamma \left( {1 + \gamma  + j\gamma } \right)\Gamma \left( {\alpha  + j\gamma } \right)\Gamma \left( {\beta  + j\gamma } \right)}}{{\Gamma \left( {1 + \gamma } \right)\Gamma \left( {\alpha  + \beta  + \left( {k + j - 1} \right)\gamma } \right)}}}.
\end{split}
\end{equation}
Based on the two lemmas, we could solve the volume calculation problem with the two defined distances.

\subsubsection{The Projective-F Norm Distance}
first, we consider the case of $\delta\leq 1$, then the volume of the ball $ {V_\mathbb{C}}\left( {\mathfrak{B}\left( \delta  \right)} \right)$ in $ \mathbb{G}\left( {k,n} \right)$ is written as
\ifCLASSOPTIONonecolumn
\begin{equation}
\label{f30}
p\left( {{d_{pF}} \leq \delta } \right) = M\int_{\mathbf{\Omega }} {\prod\limits_{i = 1}^k {{{\left( {{{\sin }^2}{\theta _i}} \right)}^{n - 2k}}} \prod\limits_{1 \leq i < j}^k {{{\left( {{{\sin }^2}{\theta _i} - {{\sin }^2}{\theta _j}} \right)}^2}} d{{\sin }^2}{\theta _1} \cdots d{{\sin }^2}{\theta _k}}.
\end{equation}
\else
\begin{align}
\label{f30}
p\left( {{d_{pF}} \leq \delta } \right) & = M\int_{\mathbf{\Omega }} \nonumber \prod\limits_{i = 1}^k {{{\left( {{{\sin }^2}{\theta _i}} \right)}^{n - 2k}}} \\ 
& \,\,\, \times \prod\limits_{1 \leq i < j}^k {{{\left( {{{\sin }^2}{\theta _i} - {{\sin }^2}{\theta _j}} \right)}^2}} d{{\sin }^2}{\theta _1} \cdots d{{\sin }^2}{\theta _k}.
\end{align}
\fi
The integration is taken over the domain in (\ref{f11}). Notice that the polynomial to be integrated is is symmetric and there are $k!$ permutaions of $ {\sin ^2}{\theta _1}, \cdots ,{\sin ^2}{\theta _k}$, which means that (\ref{f30}) could be transferred to the following equivalent integration problem
\ifCLASSOPTIONonecolumn
\begin{equation}
\label{f31}
p\left( {{d_{pF}} \leq \delta } \right) = \frac{M}{{k!}}\int_{{{\mathbf{\Omega }}_1}}{\prod\limits_{i = 1}^k {{{\left({{{\sin }^2}{\theta _i}} \right)}^{n - 2k}}} \prod\limits_{1 \leq i < j}^k {{{\left( {{{\sin }^2}{\theta _i} - {{\sin }^2}{\theta _j}} \right)}^2}} d{{\sin }^2}{\theta _1} \cdots d{{\sin }^2}{\theta _k}}
\end{equation}
\else
\begin{align}
\label{f31}
p\left( {{d_{pF}} \leq \delta } \right) & = \frac{M}{{k!}}\int_{{{\mathbf{\Omega }}_1}}\prod\limits_{i = 1}^k {{{\left({{{\sin }^2}{\theta _i}} \right)}^{n - 2k}}} \nonumber \\ 
& \,\,\, \times \prod\limits_{1 \leq i < j}^k {{{\left( {{{\sin }^2}{\theta _i} - {{\sin }^2}{\theta _j}} \right)}^2}} d{{\sin }^2}{\theta _1} \cdots d{{\sin }^2}{\theta _k}
\end{align}
\fi
where the new domain ${{\mathbf{\Omega }}_1}$ is ${\mathbf{\Omega _1}} = \{ \left. \sum\limits_{i = 1}^k {\sin }^2\theta _i  \leq \delta ^2 \right| 0 \leq {{\sin }^2}{\theta _i} \leq 1,i = 1, \cdots ,k \}$. Substituting the variables ${\sin ^2}{\theta _i}$, $i = 1, \cdots ,k$, with $ \delta {x_i} = {\sin ^2}{\theta _i}$, (\ref{f31}) is further transferred to 
\ifCLASSOPTIONonecolumn
\begin{equation}
\label{f32}
p\left( {{d_{pF}} \leq \delta } \right) = \frac{{M{\delta ^{2nk - {2k^2}}}}}{{k!}}\int_{{{\mathbf{\Omega }}_2}} {\prod\limits_{i = 1}^k {{x_i}^{n - 2k}} \prod\limits_{1 \leq i < j \leq k} {{{\left( {{x_i} - {x_j}} \right)}^2}} d{x_1} \cdots d{x_k}}
\end{equation}
\else
\begin{align}
\label{f32}
p\left( {d_{pF}} \leq \delta  \right) & = \frac{M\delta ^{2nk - {2k^2}}}{k!}\int_{\mathbf{\Omega }_2} \prod\limits_{i = 1}^k {x_i}^{n - 2k} \nonumber \\ 
& \,\,\, \times \prod\limits_{1 \leq i < j \leq k} {{\left( {{x_i} - {x_j}} \right)}^2} d{x_1} \cdots d{x_k}
\end{align}
\fi
where the integration is taken over $ {{\mathbf{\Omega }}_2} = \{ \sum\limits_{i = 1}^k {{x_i}}  \leq 1,0 \leq {x_i} \leq 1,i=1,\cdot,k \}$. According to Lemma 2, we get the final formula for the case of $\delta \leq 1$ as
\ifCLASSOPTIONonecolumn
\begin{equation}
\label{f33}
p\left( {{d_{pF}} \leq \delta } \right) = \frac{{M{\delta ^{2kn - 2{k^2}}}}}{{k!\Gamma \left( {nk - {k^2} + 1} \right)}}\prod\limits_{i = 0}^{k - 1} {\Gamma \left( {n - 2k + 1 + i} \right)\Gamma \left( {2 + i} \right)} 
\end{equation}
\else
\begin{align}
\label{f33}
p\left( {{d_{pF}} \leq \delta } \right) & = \frac{{M{\delta ^{2kn - 2{k^2}}}}}{{k!\Gamma \left( {nk - {k^2} + 1} \right)}} \nonumber \\ & \,\,\, \times \prod\limits_{i = 0}^{k - 1} {\Gamma \left( {n - 2k + 1 + i} \right)\Gamma \left( {2 + i} \right)} 
\end{align}
\fi
which is further simplified to
\begin{equation}
\label{f34}
p\left( {{d_{pF}} \leq \delta } \right) = {\delta ^{2kn - 2{k^2}}}\frac{{C{\Gamma _k}\left( n \right)}}{{\Gamma \left( {nk - {k^2} + 1} \right)C{\Gamma _k}\left( k \right)}}.
\end{equation}
For the case of $\delta>1 $, the domain of integration is a complicated polyhedron. In order to obtain the precise result, the integration domain has to be divided into $\left\lceil \delta  \right\rceil $ sections and calculated in each section respectively. Even so, it is still hard to get an analytic solution. Fortunately, the requirement of the volume formula of hyperball ${{\mathfrak{B}_{k,n}}\left( \delta  \right)}$ for $\delta  \geq 1$ when $k \geq 3$ and $n \geq 2k$  seldom occures, and if needed, the formula in reference [8] is accurate enough for practical applications. Therefore, we cease to study this problem further in this paper.

\subsubsection{The Projective-2 Norm Distance}
according to the projective-2 norm distance, the volume of $\delta$-ball in $ \mathbb{G}\left( {k,n} \right)$ is calculated as the following integration
\ifCLASSOPTIONonecolumn
\begin{equation}
\label{f35}
p\left( {{d_{p2}} \leq \delta } \right) = M\int_{\mathbf{\Omega }} {\prod\limits_{i = 1}^k {{{\left( {{{\sin }^2}{\theta _i}} \right)}^{n - 2k}}} \prod\limits_{1 \leq i < j}^k {{{\left( {{{\sin }^2}{\theta _i} - {{\sin }^2}{\theta _j}} \right)}^2}} d{{\sin }^2}{\theta _1} \cdots d{{\sin }^2}{\theta _k}} 
\end{equation}
\else
\begin{align}
\label{f35}
p\left( {{d_{p2}} \leq \delta } \right) & = M\int_{\mathbf{\Omega }} \prod\limits_{i = 1}^k {{{\left( {{{\sin }^2}{\theta _i}} \right)}^{n - 2k}}} \nonumber \\ & \,\,\, \times \prod\limits_{1 \leq i < j}^k {{{\left( {{{\sin }^2}{\theta _i} - {{\sin }^2}{\theta _j}} \right)}^2}} d{{\sin }^2}{\theta _1} \cdots d{{\sin }^2}{\theta _k} 
\end{align}
\fi
where ${\mathbf{\Omega }}$ is defined in (\ref{f12}). Again, using the property of symmetric polynomial and substituting the variable ${\sin ^2}{\theta _i}$ with ${\delta ^2}{x_i}$, $i = 1, \cdots ,k$, the above integration is transformed to
\ifCLASSOPTIONonecolumn
\begin{equation}
\label{f36}
p\left( {{d_{p2}} \leq \delta } \right) = \frac{{M{\delta ^{2nk - {k^2}}}}}{{k!}}\int_{{{\mathbf{\Omega }}_2}} {\prod\limits_{i = 1}^k {{x_i}^{n - 2k}} \prod\limits_{1 \leq i < j \leq k} {{{\left( {{x_i} - {x_j}} \right)}^2}} d{x_1} \cdots d{x_k}} 
\end{equation}
\else
\begin{align}
\label{f36}
p\left( {{d_{p2}} \leq \delta } \right) & = \frac{{M{\delta ^{2nk - {k^2}}}}}{{k!}}\int_{{{\mathbf{\Omega }}_2}} \prod\limits_{i = 1}^k {{x_i}^{n - 2k}} \nonumber \\ & \,\,\, \times \prod\limits_{1 \leq i < j \leq k} {{{\left( {{x_i} - {x_j}} \right)}^2}} d{x_1} \cdots d{x_k} 
\end{align}
\fi
where the integral is taken over ${{\mathbf{\Omega }}_2} = \{ 0 \leq {x_i} \leq 1,i = 1, \cdots ,k \}$. Using Lemma 1 and simplifying the results, we get
\begin{equation}
\label{f37}
p\left( {{d_{p2}} \leq \delta } \right) = {\delta ^{2kn - 2{k^2}}}.
\end{equation}
This means ${\bar V_{pF}}\left( {{\mathfrak{B}_{k,n}}\left( \delta  \right)} \right) = {\delta ^{2kn - 2{k^2}}}$.

\subsection{Case of $ n < 2k$ }
When $ n < 2k$, for any two points ${{\mathbf{V}}_1},{{\mathbf{V}}_2} \in \mathbb{G}\left( {k,n} \right)$, let 
\begin{equation}
\label{f38}
{{\mathbf{V}}_1} = \left[ {\begin{array}{*{20}{c}}
  {{\mathbf{V}}_1^1}&{{\mathbf{V}}_1^2} 
\end{array}} \right],{{\mathbf{V}}_2} = \left[ {\begin{array}{*{20}{c}}
  {{\mathbf{V}}_2^1}&{{\mathbf{V}}_2^2} 
\end{array}} \right]
\end{equation}
where ${\mathbf{V}}_1^1$ and ${\mathbf{V}}_2^1$ are the first $n-k$ columns of ${{\mathbf{V}}_1}$ and ${{\mathbf{V}}_2}$ respectively, while ${\mathbf{V}}_1^2$ and ${\mathbf{V}}_2^2$ are the last $2k-n$ columns of ${{\mathbf{V}}_1}$ and ${{\mathbf{V}}_2}$ respectively. As any two $k$-dimensional subspaces in a $n$-dimensional space have an overlapped subspace with dimension of $2k-n$, let ${\mathbf{V}}_1^2$ and ${\mathbf{V}}_2^2$ be the overlapped subspace, which means ${\mathbf{V}}_2^2 = {\mathbf{V}}_1^2{\mathbf{U}}$, where ${\mathbf{U}}$ is a $\left( {2k - n} \right) \times \left( {2k - n} \right)$ unitary matrix. Using ${\mathbf{V}}_2^{1,H}{\mathbf{V}}_2^2 = 0$, we get
\begin{equation}
\label{f39}
{\mathbf{V}}_1^H {\mathbf{V}}_2  = \left[ {\begin{array}{*{20}c}
   {{\mathbf{V}}_1^{1,H} {\mathbf{V}}_2^1 } & {{\mathbf{V}}_1^{1,H} {\mathbf{V}}_2^2 }  \\
   {{\mathbf{V}}_1^{2,H} {\mathbf{V}}_2^1 } & {{\mathbf{V}}_1^{2,H} {\mathbf{V}}_2^2 }  \\

 \end{array} } \right] = \left[ {\begin{array}{*{20}c}
   {{\mathbf{V}}_1^{1,H} {\mathbf{V}}_2^1 } & {\mathbf{0}}  \\
   {\mathbf{0}} & {\mathbf{U}}  \\
 \end{array} } \right].
\end{equation}
Let the SVD of $ {\mathbf{V}}_1^{1,H}{\mathbf{V}}_2^1$ be $ {\mathbf{V}}_1^{1,H}{\mathbf{V}}_2^1 = {{\mathbf{U}}_A}\cos {{\mathbf{\Theta _A} }}{\mathbf{V}}_A^H$, then ${\mathbf{V}}_1^{H}{\mathbf{V}}_2$ could be expressed as
\begin{equation}
\label{f40}
{\mathbf{V}}_1^H {\mathbf{V}}_2  = \left[ {\begin{array}{*{20}c}
   {{\mathbf{U}}_A } & {\mathbf{0}}  \\
   {\mathbf{0}} & {\mathbf{U}}  \\

 \end{array} } \right]\left[ {\begin{array}{*{20}c}
   {\cos {\mathbf{\Theta _A} }} & {\mathbf{0}}  \\
   {\mathbf{0}} & {\cos {\mathbf{0}}}  \\

 \end{array} } \right]\left[ {\begin{array}{*{20}c}
   {{\mathbf{V}}_A^H } & {\mathbf{0}}  \\
   {\mathbf{0}} & {\mathbf{I}}  \\
 \end{array} } \right]
\end{equation}
where $\cos {\mathbf{\Theta _A} }$ is a $\left( {n - k} \right) \times \left( {n - k} \right)$ diagonal matrix and ${\mathbf{\Theta _A} }$ is the canonical angle matrix in
$\mathbb{G}\left( {n - k,n} \right)$. According to the distance definitions, the volume of hyperball in $\mathbb{G}\left( {k,n} \right)$ is equal to that of $\mathbb{G}\left( {n - k,n} \right)$. It also means
\begin{equation}
\label{f41}
\bar V\left( {{\mathfrak{B}_{k,n}}\left( \delta  \right)} \right) = \bar V\left( {{\mathfrak{B}_{n - k,n}}\left( \delta  \right)} \right).
\end{equation}

\subsection{Numerical Simulation and Discussion}
In this section, we present some illustrative numerical results to show that the formulas provided in Section III yield very accurate results in various cases. In simulation, we generate ${10^6}$ couples of uniformly distributed points in $\mathbb{G}\left( {k,n} \right)$, and calculate the projective-F norm distance and Projective-2 norm distance of each couple of points. According to (\ref{f8}), the CDF of these distance values denotes the probability $ p\left( {d \leq \delta } \right) = {\bar V_d}\left( {{\mathfrak{B}_{k,n}}\left( \delta  \right)} \right)$. In order to generate uniformly distributed points in $\mathbb{G}\left( {k,n} \right)$, a random matrix ${\mathbf{A}} \in {\mathbb{C}^{n \times k}}$ with $\mathcal{C}\mathcal{N}\left( {0,1} \right)$ elements is generated first. Next, let the thin QR decomposition of ${\mathbf{A}}$ be ${\mathbf{A}} = {\mathbf{QR}}$ where the diagonal elements of the matrix $\mathbf{R}$ is positive (e.g., implemented by the Gram-Schmidt process), then the matrix ${\mathbf{Q}}$ is the desired matrix. Here we provide a simple explanation of this method. For more details please refer to [21]. The Ginibre ensemble $G$ consists of matrices ${\mathbf{A}} \in \mathbb{C}^{n \times k} $, whose elements $a_{ij} $ are independent and identically distributed (i.i.d.) standard normal complex random variables. Then, the probability density function of $\mathbf A$ is $ f_G \left( {\mathbf{A}} \right) = \frac{1}{{\pi ^{kn} }}\exp \left( { - {\text{tr}}\left( {{\mathbf{A}}^{\text{H}} {\mathbf{A}}} \right)} \right)$. Let the probability density function $f_G$ define a measure of $G$ as $d\mu _G \left( {\mathbf{A}} \right): = f_G \left( {\mathbf{A}} \right)d{\mathbf{A}}$. Given any ${\mathbf{A}}\in G$, we define the equivalent class
$\left[ {\mathbf{A}} \right] = \left\{ {\left. {{\mathbf{UQR}}} \right|{\mathbf{U}} \in \mathbb{U}\left( {n} \right)} \right\}$. Note that the set $ \left\{ {\left. {{\mathbf{UQ}}} \right|{\mathbf{U}} \in \mathbb{U}\left( {n} \right)} \right\}$ forms the Stiefel manifold $\mathbb{S}\mathbb{T}\left( {k,n}\right)$. Since the measure $d\mu _G $ is invariant under left-multiplication by $\mathbb{U}\left( {n} \right)$, the restriction of $d\mu_G$ to equivalent
$\left[ {\mathbf{A}} \right]$ is also left-invariant for every $ {\mathbf{A}} \in G$. According to the Harr measure theory, it means that the matrix $\mathbf{Q}$ is uniformly distributed on $\mathbb{S}\mathbb{T}\left( {k,n}\right)$ given any matrix $ {\mathbf{A}} \in G$. Therefore, the method generates uniformly distributed points in Grassmann manifold considering the equivalent relationship in Stiefel manifold. The comparisons between the simulated and calculated results are shown in Fig. 1, Fig. 2, Fig. 3, and Fig. 4 for various cases.

\begin{figure}[!t]
\centering \includegraphics[width = 0.8\linewidth]{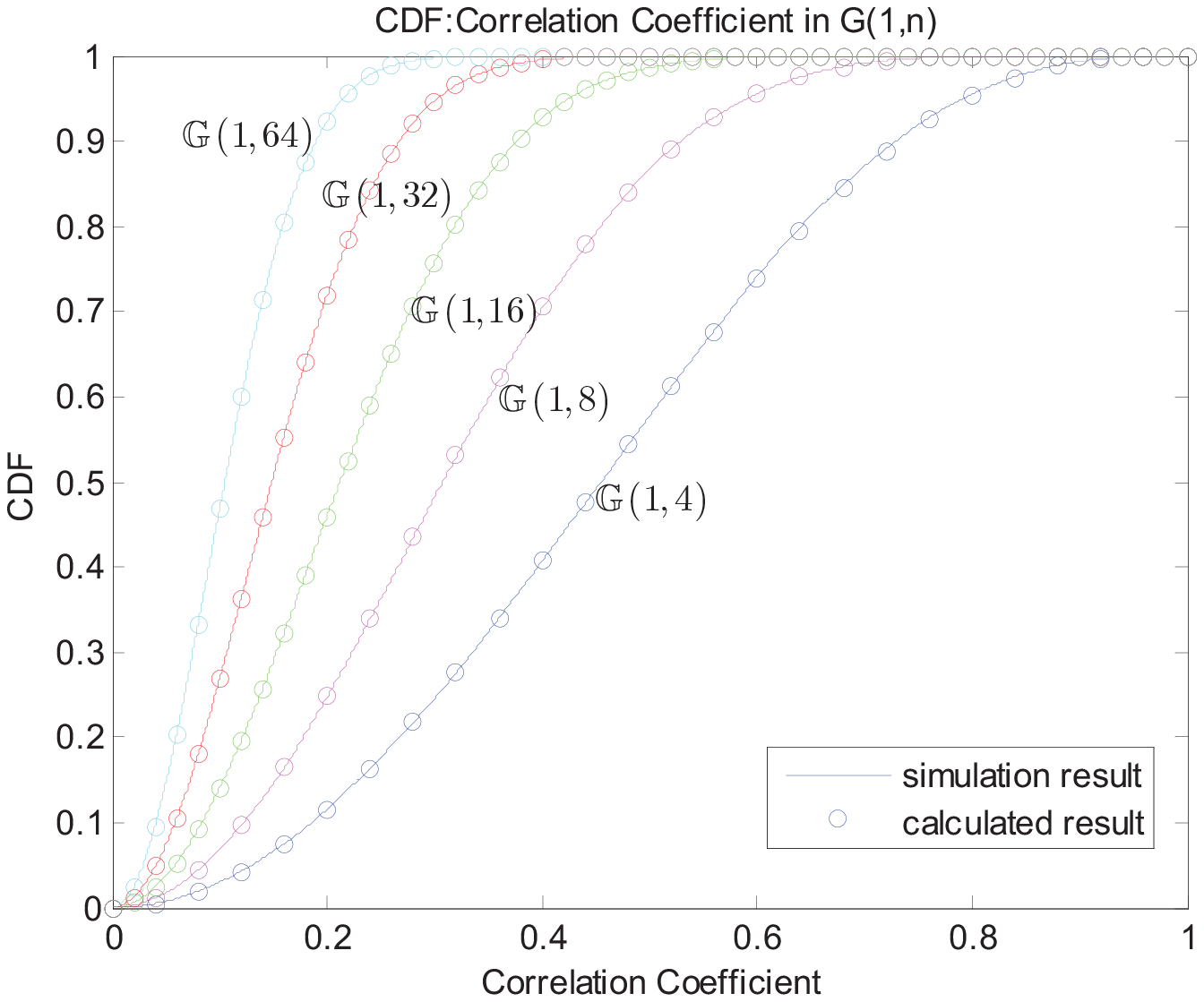}
\caption{$r\left( n \right)$ in (\ref{f15}) vs. $n$ in $\mathbb{G}\left( {1,n} \right)$}
\end{figure}

\begin{figure}[!t]
\centering \includegraphics[width = 0.8\linewidth]{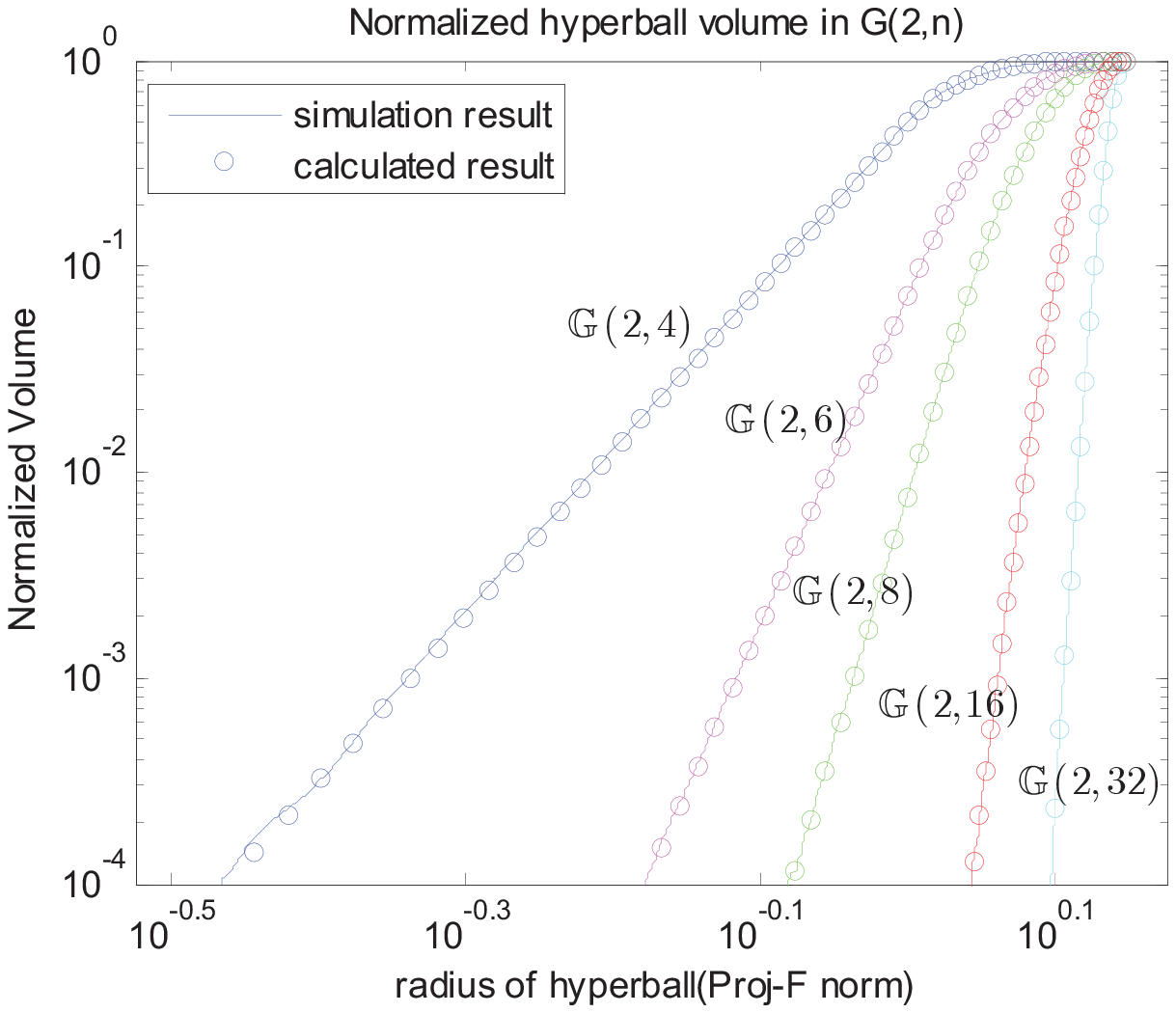}
\caption{${\bar V_{{d_{pF}}}}\left( {{\mathfrak{B}_{2,n}}\left( \delta  \right)} \right)$ vs. $n$ in $\mathbb{G}\left( {2,n} \right)$}
\end{figure}

\begin{figure}[!t]
\centering \includegraphics[width = 0.8\linewidth]{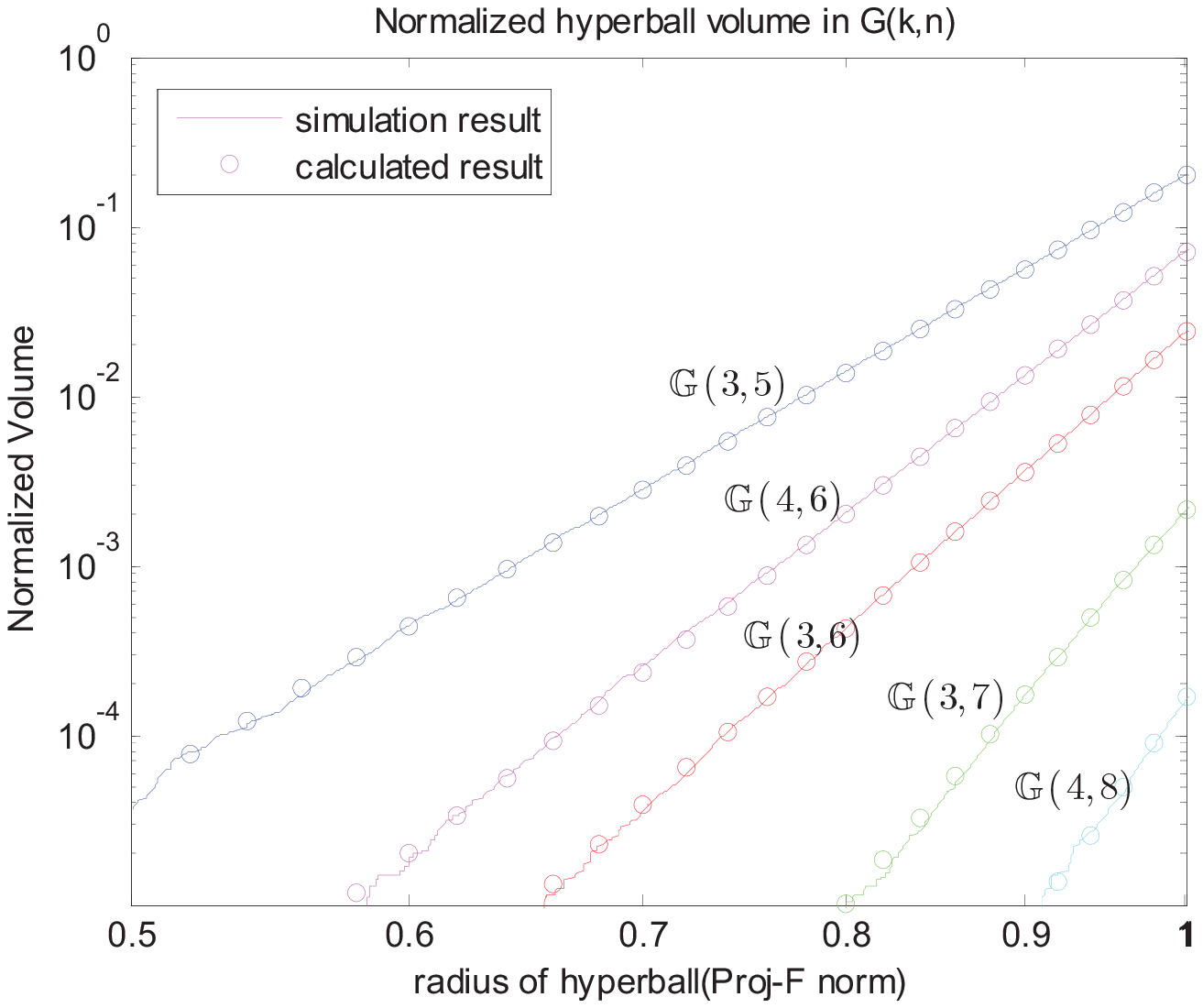}
\caption{${{\bar V}_{{d_{p2}}}}\left( {{\mathfrak{B}_{k,n}}\left( \delta  \right)} \right)$ vs. $\left( {k,n} \right)$ in $\mathbb{G}\left( {k,n} \right)$}
\end{figure}

\begin{figure}[!t]
\centering \includegraphics[width = 0.8\linewidth]{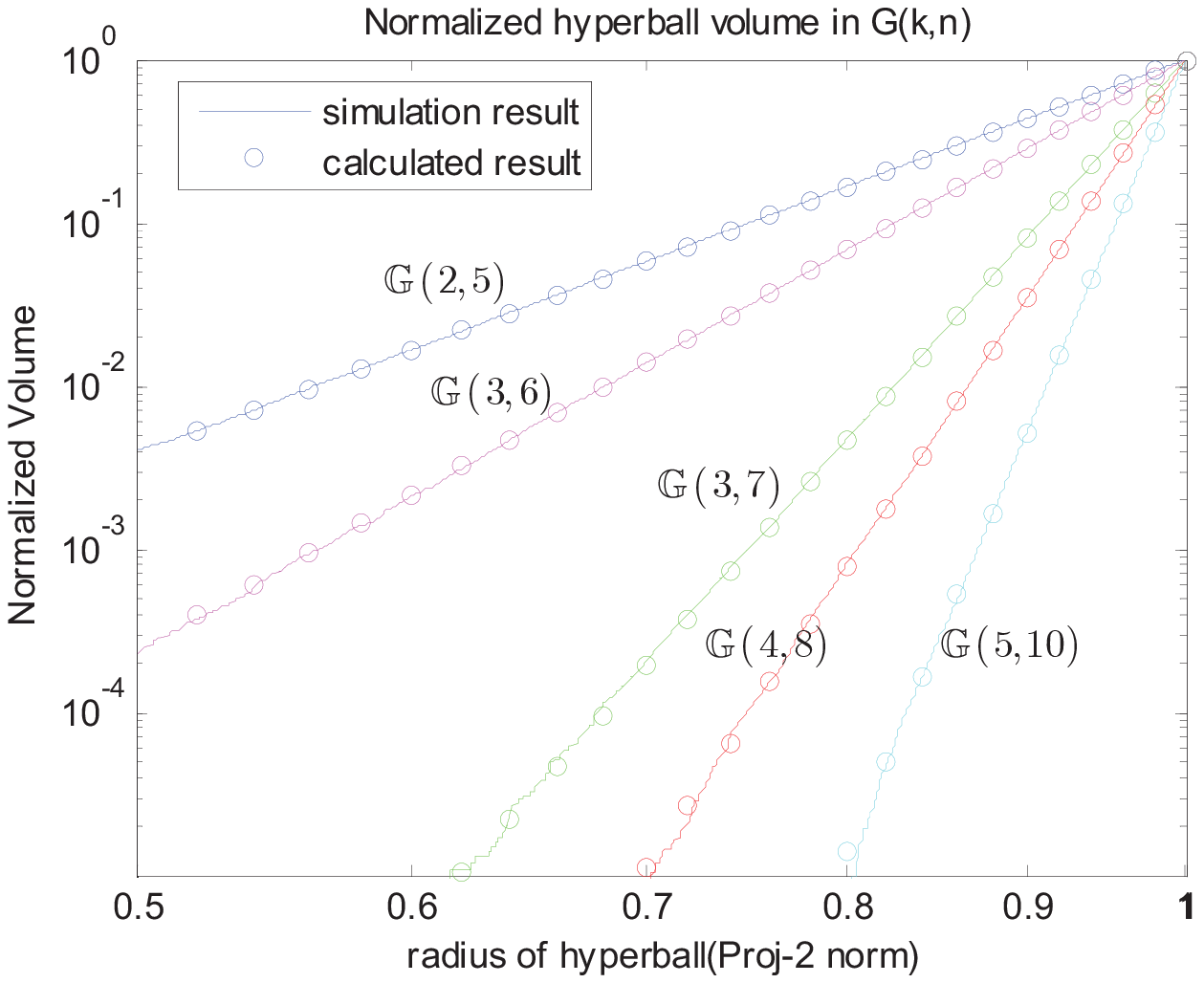}
\caption{${{\bar V}_{{d_{pF}}}}\left( {{\mathfrak{B}_{k,n}}\left( \delta  \right)} \right)$ vs. $\left( {k,n} \right)$ in $\mathbb{G}\left( {k,n} \right)$}
\end{figure}

In Fig. 1, the CDF of correlation coefficient in $\mathbb{G}\left( {1,n} \right)$ is obtained by Monte Carlo simulation (solid line) and it is compared with the corresponding values (circle) from closed form (\ref{f15}). Obviously, we can say that the simulation results match the calculated results perfectly. For $k=2$ with projective-F norm distance, the simulated and calculated volume of hyperball with radius $\delta$ are compared in Fig. 2, while for $k > 2$ and $\delta\leq 1$, similar comparisons are shown in Fig. 3. These two figures show the accuracy of (\ref{f18}), (\ref{f34}), and (\ref{f41}). In Fig. 4, the calculated values of formulas (\ref{f37}) and (\ref{f41}) are verified by simulation results for various $k$ and $n$. Again, they match each other almost perfectly. All the numerical results indicate that the closed-form formulas in this paper are accurate. Furthermore, as ${\bar V_d}\left( {{\mathfrak{B}_{k,n}}\left( \delta  \right)} \right)$ is an exponential function of the dimension of complex vector space $n$ for given $\delta$ and $k$, the volume of the hyperball with radius $\delta$ decreases very fast as $n$ grows, which means that the distance of any two $k$-dimensional subspaces in ${\mathbb{C}^n}$ approaches to the maximum value rapidly as $n$ grows. In other words,  the probability of any two $k$-dimensional subspaces being mutually orthogonal approaches to 1 very fast. This provdes a foundation for large-scale MU-MIMO communication systems where the eigenspace of the wireless channel matrix is a point in $\mathbb{G}\left( {k,n} \right)$.

\section{Application in Large-Scale MIMO Communication Systems}
In this section, the formulas in Section IV are applied to analyze the SINRs of MU-MMO with non-ideal CSI in large-scale MIMO systems. Firstly, an error model of CSI is constructed, then an
approximation expression of SINR of each user is derived. After that, the expectation of SINR is obtained based on the probability density function of correlation coefficient between any two points in complex
Grassmann manifold. This expectation can be used to predict each user's data transmission rate in MU-MIMO systems with non-ideal CSI, and to predict the total data transmission rate in MU-MIMO
systems. Finally, simulation results are provided to verify the validity of the approximation expression.

\subsection{System Model}
Considering a wireless communication system where a BS serves $K$ users. Suppose that the BS has $N$ transmit antennas and each user has one receive antenna. In this system, the BS chooses the transmission mode adaptively according to the information of the $K$ users when transmitting data. The adaptive transmission mode includes switching between SU-MIMO and MU-MIMO, and changing the the selection of users to serve in MU-MIMO mode. In this paper, we assume that the BS only has CQI of each user in SU-MIMO mode as ${\text{SIN}}{{\text{R}}^{{\text{SU}}}}$, which is the case in practical systems, e.g., 3GPP LTE/LTE-A. Let ${{\mathbf{h}}_i} = \left[ {\begin{array}{*{20}{c}}{{h_{i,1}}}& \cdots &{{h_{i,N}}} \end{array}} \right]$, $i = 1, \cdots ,K$, be the channel vector of the $i$th user with $\mathcal{C}\mathcal{N}\left( {0,1} \right)$ elements, then the eigenvector space of ${{\mathbf{h}}_i}$ is ${{\mathbf{u}}_i} = {{{{\mathbf{h}}_i}} \mathord{\left/{\vphantom {{{{\mathbf{h}}_i}} {{{\left\| {{{\mathbf{h}}_i}} \right\|}_2}}}} \right.
 \kern-\nulldelimiterspace} {{{\left\| {{{\mathbf{h}}_i}} \right\|}_2}}}$. Because of system errors, e.g., limited feedback bandwidth or measurement errors, the BS obtains the CSI as ${{\mathbf{v}}_i}$ instead of 
${{\mathbf{u}}_i}$, and the relation between them could be modeled as
\begin{equation}
\label{f42}
{{\mathbf{u}}_i} = \alpha_i {e^{j{\theta _i}}}{{\mathbf{v}}_i} + \sqrt {1 - {\alpha ^2}} {e^{j{w_i}}}{\mathbf{v}}_i^ \bot ,i = 1, \cdots ,K
\end{equation}
where $\alpha_i $ is the correlation coefficient between ${{\mathbf{u}}_i}$ and ${{\mathbf{v}}_i}$ defined as $\left| {{{\mathbf{u}}_i}{\mathbf{v}}_i^H} \right|$. Since only the $i$th user is studied in this paper, the $\alpha$ is used to replace $\alpha_i$ for simplification. Although $\alpha $ satisfies the constraint $0 \ll \alpha  < 1$, it should be much closer to 1 than to 0 so that the MU-MIMO system can work in practice. In a TDD system, by making use of the channel reciprocity, the BS obtains the downlink CSI through the pilot sent by users in the uplink, hence $\alpha$ is mainly determined by the SINR of each user in the uplink, and it is different for each user. The symbol ${\mathbf{v}}_i^ \bot $ represents an unit vector in the null space of ${{\mathbf{v}}_i}$. Obviously, ${{\mathbf{u}}_i}$, ${{\mathbf{v}}_i}$, and ${\mathbf{v}}_i^ \bot $ are elements in $\mathbb{G}\left( {1,N} \right)$. At the BS side, the data of $K$ users transmitted on the same time-frequency resource in MU-MIMO systems are mapped onto the $N$ antennas by the precoding matrix ${\mathbf{W}} = \left[ {\begin{array}{*{20}c}{{\mathbf{w}}_1 } &  \cdots  & {{\mathbf{w}}_K }\\\end{array} } \right]$, where ${\mathbf{w}}_i $ is the $N$-dimensional precoding vector belonging to the $i$th user. Then, the SINR of the $i$th user with non-ideal CSI could be written as
\begin{equation}
\label{f43}
{\mathrm{SINR}}_i^{{\mathrm{MU,non - ideal}}}  = \frac{{\left\| {{\bf{h}}_i {\bf{w}}_i } \right\|^2 }}
{{\sum\limits_{\scriptstyle j = 1 \hfill \atop 
  \scriptstyle j \ne i \hfill} ^K {\left\| {{\bf{h}}_i {\bf{w}}_j } \right\|^2 }  + \sigma _{NI}^2 }},i = 1, \cdots ,K
\end{equation}
where $\sigma _{NI}^2$ denotes the power of noise plus interference from neighboring cells. Next, we analyze the SINR based on two general precoding methods: Conjugate Beamforming (CB) and ZF.

\subsection{Conjugate Beamforming}

\subsubsection{Estimation of SINR}
for CB, the precoding vector of the $i$th user is 
\begin{equation}
\label{f44}
{\bf{w}}_i^{CB}  = c_i {\bf{h}}_i^H 
\end{equation}
where $c_i$ is a scale factor which denotes the power allocated to the $i$th user. In this paper, we assume that the total transmit power is $P$, and it is equally distributed among users. 
Hence, $ c_i  = \frac{{\sqrt P }}{{\sqrt K \left\| {{\bf{h}}_i } \right\|}}$, and ${\bf{w}}_i^{CB}  = \sqrt {\frac{P}{K}} {\bf{v}}_i $. Then, the SINR of the $i$th user is
\ifCLASSOPTIONonecolumn
\begin{equation}
\label{f45}
{\mathrm{SINR}}_i^{{\mathrm{CB,non - ideal}}}  = \frac{{P\left| {{\bf{h}}_i {\bf{v}}_i^H } \right|^2 }}
{{\sum\limits_{\scriptstyle j = 1 \hfill \atop 
  \scriptstyle j \ne i \hfill} ^K {P\left| {{\bf{h}}_i {\bf{v}}_j^H } \right|^2 }  + K\sigma _{NI}^2 }} = \frac{{\left| {{\bf{u}}_i {\bf{v}}_i^H } \right|^2 }}
{{\sum\limits_{\scriptstyle j = 1 \hfill \atop 
  \scriptstyle j \ne i \hfill} ^K {\left| {{\bf{u}}_i {\bf{v}}_j^H } \right|^2 }  + K\gamma }}
\end{equation}
\else
\begin{align}
\label{f45}
{\mathrm{SINR}}_i^{{\mathrm{CB,non - ideal}}}  & = \frac{{P\left| {{\bf{h}}_i {\bf{v}}_i^H } \right|^2 }}
{{\sum\limits_{\scriptstyle j = 1 \hfill \atop 
  \scriptstyle j \ne i \hfill} ^K {P\left| {{\bf{h}}_i {\bf{v}}_j^H } \right|^2 }  + K\sigma _{NI}^2 }} \nonumber \\ & = \frac{{\left| {{\bf{u}}_i {\bf{v}}_i^H } \right|^2 }}
{{\sum\limits_{\scriptstyle j = 1 \hfill \atop 
  \scriptstyle j \ne i \hfill} ^K {\left| {{\bf{u}}_i {\bf{v}}_j^H } \right|^2 }  + K\gamma }}
\end{align}
\fi
where $\gamma  = \frac{{\sigma _{NI}^2 }}{{P\left\| {{\bf{h}}_i } \right\|^2 }} = \frac{1}{{{\text{SINR}}^{{\text{SU}}} }}$. Substituting (\ref{f42}) into (\ref{f45}) and making use of $
{\bf{v}}_i^ \bot  {\bf{v}}_i^H  = 0$, it becomes 
\ifCLASSOPTIONonecolumn
\begin{equation}
\label{f46}
\begin{split}
 {\mathrm{SINR}}_i^{{\mathrm{CB,non - ideal}}}  &=  \frac{{\alpha ^2 }}
{{\sum\limits_{\scriptstyle j = 1 \hfill \atop 
  \scriptstyle j \ne i \hfill} ^K {\left| {\alpha e{}^{j\theta _i }{\bf{v}}_i {\bf{v}}_j^H  + \sqrt {1 - \alpha ^2 } e{}^{j\theta _i }{\bf{v}}_i^ \bot  {\bf{v}}_j^H } \right|^2 }  + K\gamma }}  \\
  & {\mathrm{                  }}\mathop  \approx \limits^{ 1} \frac{{\alpha ^2 }}
{{\sum\limits_{\scriptstyle j = 1 \hfill \atop 
  \scriptstyle j \ne i \hfill} ^K {\left( {\left| {\alpha {\bf{v}}_i {\bf{v}}_j^H } \right|^2  + \left( {1 - \alpha ^2 } \right)\left| {{\bf{v}}_i^ \bot  {\bf{v}}_j^H } \right|^2 } \right)}  + K\gamma }}  \\ 
  & {\text{                  }}\mathop  \approx \limits^{2} \frac{{\alpha ^2 }}
{{\sum\limits_{\scriptstyle j = 1 \hfill \atop 
  \scriptstyle j \ne i \hfill} ^K {\left| {{\bf{v}}_i {\bf{v}}_j^H } \right|^2 }  + K\gamma }}.
\end{split}
\end{equation}
\else
\begin{align}
\label{f46}
 & {\mathrm{SINR}}_i^{{\mathrm{CB,non - ideal}}} \nonumber \\ 
& =  \frac{{\alpha ^2 }}{{\sum\limits_{\scriptstyle j = 1 \hfill \atop  \scriptstyle j \ne i \hfill} ^K {\left| {\alpha e{}^{j\theta _i }{\bf{v}}_i {\bf{v}}_j^H  + \sqrt {1 - \alpha ^2 } e{}^{j\theta _i }{\bf{v}}_i^ \bot  {\bf{v}}_j^H } \right|^2 }  + K\gamma }} \nonumber \\ 
& {\mathrm{                  }}\mathop  \approx \limits^{ 1} \frac{{\alpha ^2 }}
{{\sum\limits_{\scriptstyle j = 1 \hfill \atop 
  \scriptstyle j \ne i \hfill} ^K {\left( {\left| {\alpha {\bf{v}}_i {\bf{v}}_j^H } \right|^2  + \left( {1 - \alpha ^2 } \right)\left| {{\bf{v}}_i^ \bot  {\bf{v}}_j^H } \right|^2 } \right)}  + K\gamma }}  \nonumber \\ 
& {\text{                  }}\mathop  \approx \limits^{2} \frac{{\alpha ^2 }}
{{\sum\limits_{\scriptstyle j = 1 \hfill \atop 
  \scriptstyle j \ne i \hfill} ^K {\left| {{\bf{v}}_i {\bf{v}}_j^H } \right|^2 }  + K\gamma }}.
\end{align}
\fi
In (46), as $N$ is large, the correlation coefficients $\left| {{\bf{v}}_i {\bf{v}}_j^H } \right|$ and $\left| {{\bf{v}}_i^ \bot  {\bf{v}}_j^H } \right|$ are almost the same and close to zero, hence ignoring the term ${\bf{v}}_i {\bf{v}}_j^H {\bf{v}}_j {\bf{v}}_i^{ \bot ,H}$ in approaximation 1 and replacing $\left| {{\bf{v}}_i^ \bot  {\bf{v}}_j^H } \right|$ with $\left| {{\bf{v}}_i {\bf{v}}_j^H } \right|$ in approaximation 2 would cause little error to the estimated SINR. Actually, when the CSI is ideal, the approximation (\ref{f46}) is equal to the accurate SINR.

When $K$ is large, the term $\sum\limits_{\scriptstyle j = 1 \hfill \atop \scriptstyle j \ne i \hfill} ^K {\left| {{\bf{v}}_i {\bf{v}}_j^H } \right|^2 } $ is close to $( K - 1 ){\mathbb E}[ \left| {{\bf{v}}_i {\bf{v}}_j^H } \right|^2  ]$. Let $x = \left| {{\bf{v}}_i {\bf{v}}_j^H } \right|,{\bf{v}}_i ,{\bf{v}}_j  \in {\mathbb G}\left( {1,N} \right)$, then it is a random variable denoting the correlation coefficient between any two points in ${\mathbb G}\left( {1,N} \right)$. Hence, the probability function of $x$ could be derived from (\ref{f15}) as
\begin{equation}
\label{f47}
f\left( x \right) = 2\left( {N - 1} \right)x\left( {1 - x^2 } \right)^{N - 2} ,0 \leq x \leq 1.
\end{equation}
Using the definition of beta function, ${\mathbb E}\left[ {x^2 } \right]$ is calculated as
\begin{equation}
\label{f48}
{\mathbb E}\left[ {x^2 } \right] = \int_0^1 {x^2 f\left( x \right)dx}  = \left( {N - 1} \right)B\left( {2,N - 1} \right) = \frac{1}{N}.
\end{equation}
Therefore, when $K$ is large, the SINR could be further simplified to
\begin{equation}
\label{f49}
{\mathrm{SINR}}_i^{{\mathrm{CB,non - ideal}}}  \approx \frac{{\alpha ^2 N}}{{K + KN\gamma  - 1}}.
\end{equation}
With (\ref{f49}), we can predict the capacity of large-scale MU-MIMO systems when CB is used.

\subsubsection{Discussion about (\ref{f46}) and (\ref{f49})}
the estimation error of (\ref{f46}) is mainly caused by ignoring the cross term $\alpha \sqrt {1 - \alpha ^2 } \sum\limits_{\scriptstyle j = 1 \hfill \atop \scriptstyle j \ne i \hfill} ^K {{\bf{v}}_i {\bf{v}}_j^H {\bf{v}}_j {\bf{v}}_i^ \bot  }$. As ${\bf{v}}_i {\bf{v}}_j^H $ and ${\bf{v}}_j {\bf{v}}_i^ \bot$ are independent variables, the cross term approaches to the fixed value $\alpha \sqrt {1 - \alpha ^2 } K{\mathbb E}\left[ {{\bf{v}}_i {\bf{v}}_j^H {\bf{v}}_j^{} {\bf{v}}_i^ \bot  } \right]$ gradually as $K$ grows, which means that the standard derivation (std.) of the estimation error decreases when $K$ grows. On the other hand, the values of ${\bf{v}}_i {\bf{v}}_j^H $ and  ${\bf{v}}_j {\bf{v}}_i^ \bot$ are closer and closer to 0 when $N$ increases, thus the std. of estimation error decreases correspondingly. Furthermore, as $\alpha$ is close to 1, $\alpha \sqrt {1 - \alpha ^2 }$ decreases rapidly when $\alpha$ increases, hence as the accuracy of CQI estimation increases, the std. of estimation error decreases. For (49), it is mostly affected by the value of $K$ according to its derivation process, hence increasing $K$ will decrease the estimation error of (\ref{f49}).

\subsubsection{Numerical Results}

\begin{figure}[!t]
\centering \includegraphics[width = 0.8\linewidth]{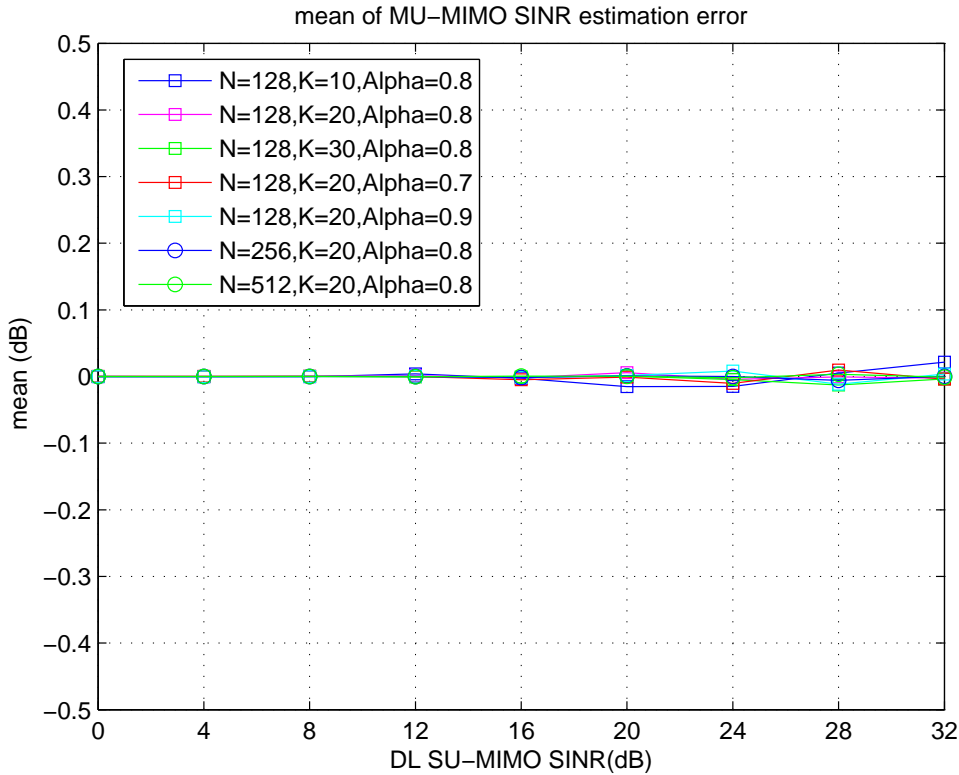}
\caption{Mean of the estimation error of (\ref{f46})}
\end{figure}

\begin{figure}[!t]
\centering \includegraphics[width = 0.8\linewidth]{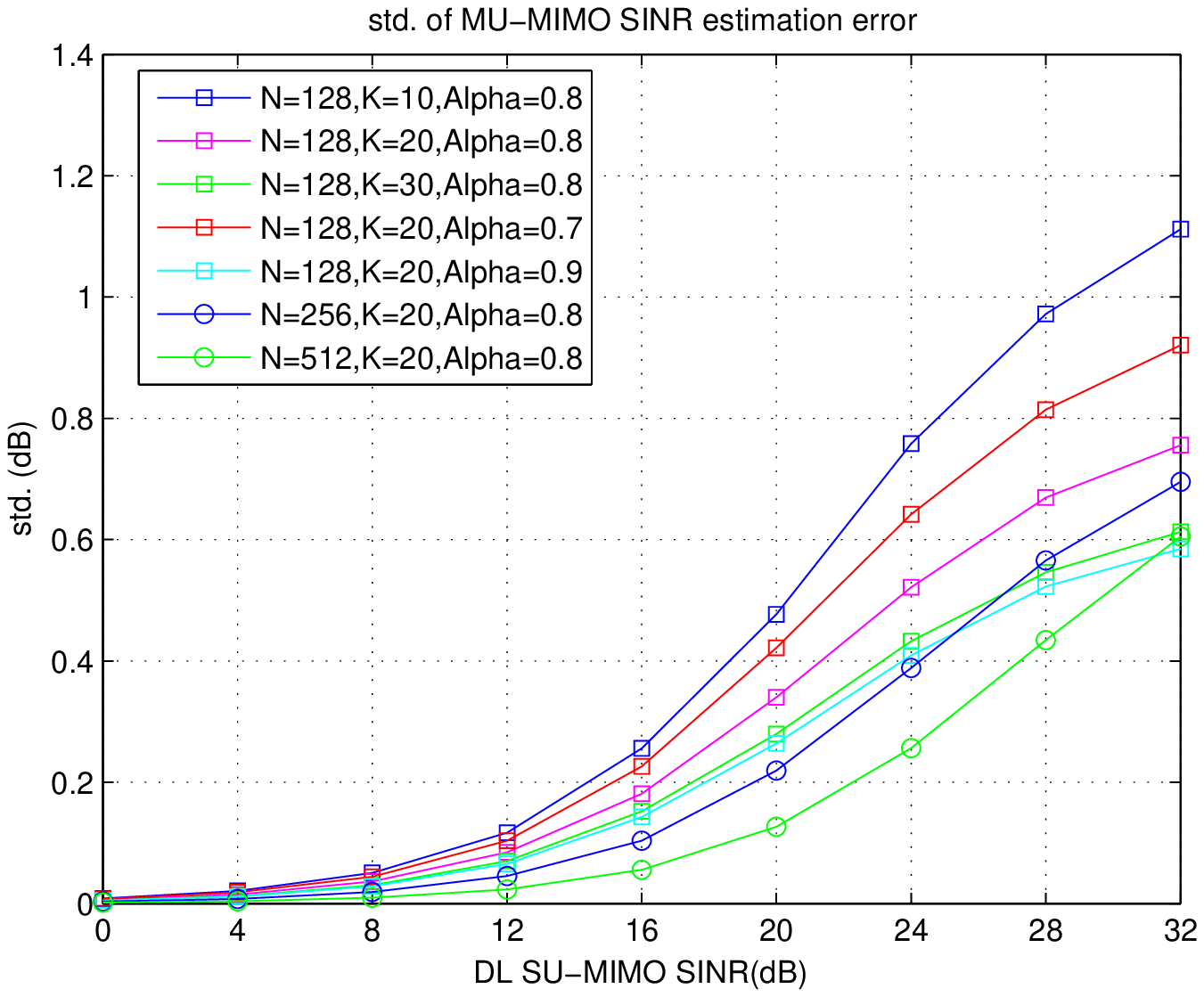}
 \caption{Std. of the estimation error of (\ref{f46})}
\end{figure}

\begin{figure}[!t]
\centering \includegraphics[width = 0.8\linewidth]{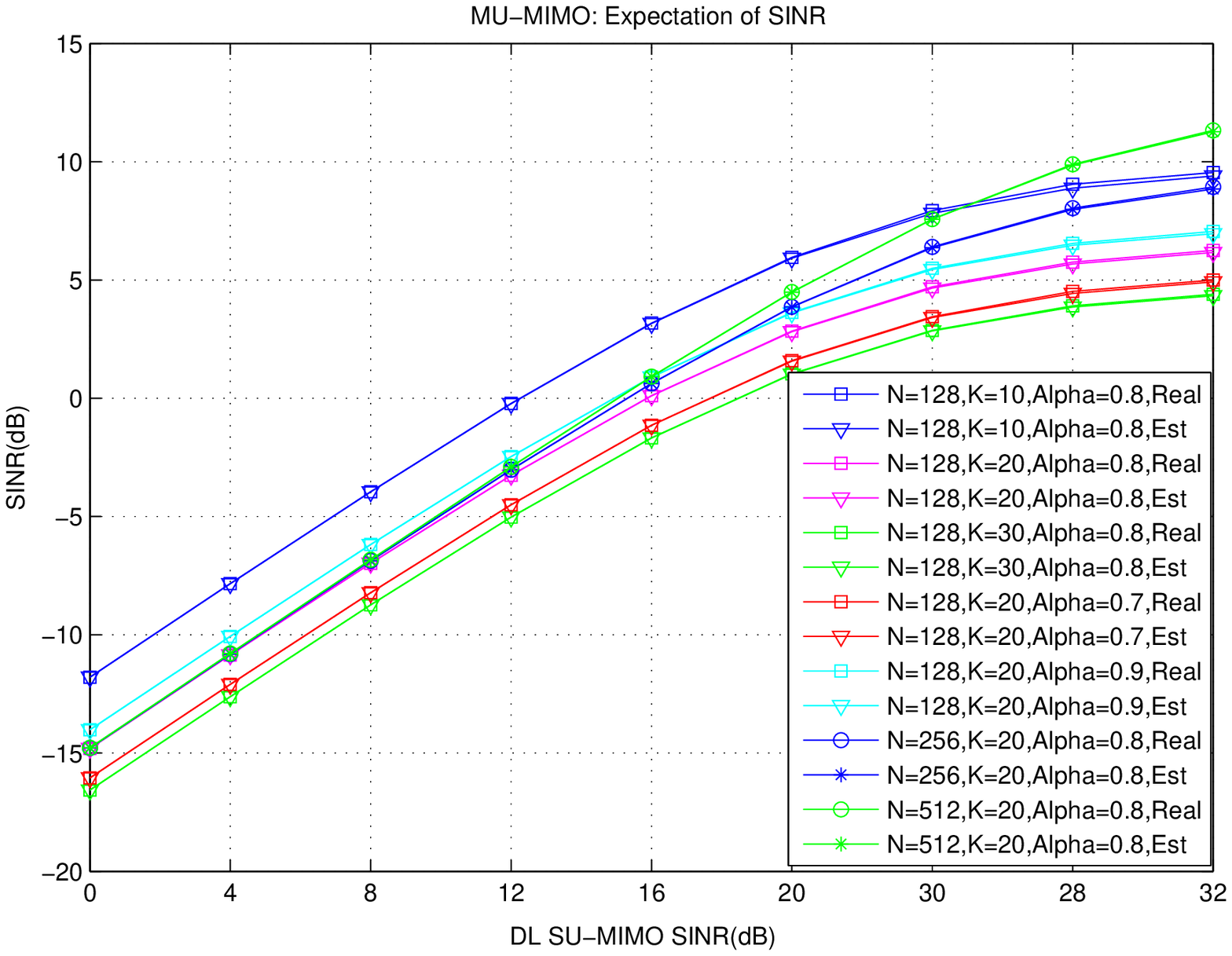}
\caption{Comparison between (\ref{f49}) and expectation of SINR}
\end{figure}

in order to verify the validity of (46), we provide some numerical results in Fig. 5 and Fig. 6, where the mean and std. of estimation errors are used to measure the accuracy of (\ref{f46}). As $\alpha$ is mainly determined by the uplink SINRs of users, we consider three typical SINR values of 0, 3, and 6 dB, which correspond to the $\alpha$ values of 0.7, 0.8, and 0.9 respectively according to (\ref{f42}). If the uplink SINR of a user is less than 0 dB, we consider it not suitable for MU-MIMO. In Fig. 5, we can see that the mean of estimation error is almost 0 regardless of the parameters $N$, $K$, and $\alpha$, which means that (\ref{f46}) provides an unbiased estimation of SINR. The std. of estimation error, which is affected by $N$, $K$, and $\alpha$, is shown in Fig. 6, and it is consistant with the analysis in 2) above. The accuracy of (\ref{f49}) is shown in Fig. 7, where the estimated values (Est. in Fig. 7) are compared with the real values (Real in Fig. 7). Obviously, (\ref{f49}) provides an approximation of the expectation of SINR with very high accuracy, hence it is helpful for us to carry out capacity estimation.

\subsection{Zero-Forcing}

\subsubsection{Estimation of SINR}
ZF precoding is used to remove the multi-user interference completely, where the multi-user interference channel of the $i$th user is defined as
\begin{equation}
\label{f50}
{\mathbf{\tilde H}}_i  = \left[ {\begin{array}{*{20}c}
   {{\mathbf{v}}_1^T } &  \cdots  & {{\mathbf{v}}_{i - 1}^T } & {{\mathbf{v}}_{i + 1}^T } &  \cdots  & {{\mathbf{v}}_K^T }  \\
 \end{array} } \right]^T.
\end{equation}
Applying the null space projection, the precoding vector of the $i$th user could be written as
\begin{equation}
\label{f51}
{\bf{w}}_i^{{\mathrm{ZF}}}  = \sqrt {\frac{P}
{K}} \frac{{\left( {{\bf{I}} - {\bf{\tilde H}}_i^H \left( {{\bf{\tilde H}}_i {\bf{\tilde H}}_i^H } \right)^{ - 1} {\bf{\tilde H}}_i } \right){\bf{v}}_i^H }}
{{\left\| {\left( {{\bf{I}} - {\bf{\tilde H}}_i^H \left( {{\bf{\tilde H}}_i {\bf{\tilde H}}_i^H } \right)^{ - 1} {\bf{\tilde H}}_i } \right){\bf{v}}_i^H } \right\|}}.
\end{equation}
Although ZF could remove the multi-user interference, it would reduce the sum of channel capacity if the grouped users for MU-MIMO have high mutual channel correlation. Therefore, the BS should group the users whose mutual correlation coefficients $\left| {{\bf{v}}_i {\bf{v}}_j^H } \right|,1 \leq i,j \leq K,i \ne j$, are lower than a predefined threshold. Since ${\bf{v}}_i  \in {\mathbb G}\left( {1,N} \right),i = 1, \cdots ,K$, ${\bf{\tilde H}}_i {\bf{\tilde H}}_i^H $ approaches to ${\bf{I}}_{K - 1}$ when $N$ becomes large according to the conclusion in Section III. Hence, it is reasonable to replace $ \left( {{\bf{\tilde H}}_i {\bf{\tilde H}}_i^H } \right)^{ - 1} $ with $ {\bf{I}}_{K - 1} $ to simplify the analysis in this paper, and (\ref{f51}) becomes
\begin{equation}
\label{f52}
{\bf{w}}_i^{{\mathrm{ZF}}}  = \frac{{\left( {{\bf{I}} - {\bf{\tilde H}}_i^H {\bf{\tilde H}}_i } \right){\bf{v}}_i^H }}
{{\left\| {\left( {{\bf{I}} - {\bf{\tilde H}}_i^H {\bf{\tilde H}}_i } \right){\bf{v}}_i^H } \right\|}}.
\end{equation}
Without losing generality, considering the $i$th user, substituting (\ref{f52}) into (\ref{f43}) and after derivation, it becomes 
\ifCLASSOPTIONonecolumn
\begin{equation}
\label{f53}
{\mathrm{SINR}}_i^{{\mathrm{ZF,non - ideal}}}  = \frac{{\left\| {{\bf{u}}_i \left( {{\bf{I}} - {\bf{\tilde H}}_i^H {\bf{\tilde H}}_i } \right){\bf{v}}_i^H } \right\|^2 }}
{{\left\| {{\bf{u}}_i \left( {{\bf{I}} - {\bf{\tilde H}}_i^H {\bf{\tilde H}}_i } \right){\bf{v}}_i^H } \right\|^2 }}\frac{1}
{{\sum\limits_{\scriptstyle j = 1 \hfill \atop 
  \scriptstyle j \ne i \hfill} ^K {\frac{{\left\| {{\bf{u}}_i \left( {{\bf{I}} - {\bf{\tilde H}}_j^H {\bf{\tilde H}}_j } \right){\bf{v}}_j^H } \right\|^2 }}
{{\left\| {\left( {{\bf{I}} - {\bf{\tilde H}}_j^H {\bf{\tilde H}}_j } \right){\bf{v}}_j^H } \right\|^2 }}}  + K\gamma }}.
\end{equation}
\else
\begin{align}
\label{f53}
{\mathrm{SINR}}_i^{{\mathrm{ZF,non - ideal}}} & = \frac{{\left\| {{\bf{u}}_i \left( {{\bf{I}} - {\bf{\tilde H}}_i^H {\bf{\tilde H}}_i } \right){\bf{v}}_i^H } \right\|^2 }}
{{\left\| {{\bf{u}}_i \left( {{\bf{I}} - {\bf{\tilde H}}_i^H {\bf{\tilde H}}_i } \right){\bf{v}}_i^H } \right\|^2 }} \nonumber \\ & \,\,\, \times \frac{1}
{{\sum\limits_{\scriptstyle j = 1 \hfill \atop 
  \scriptstyle j \ne i \hfill} ^K {\frac{{\left\| {{\bf{u}}_i \left( {{\bf{I}} - {\bf{\tilde H}}_j^H {\bf{\tilde H}}_j } \right){\bf{v}}_j^H } \right\|^2 }}
{{\left\| {\left( {{\bf{I}} - {\bf{\tilde H}}_j^H {\bf{\tilde H}}_j } \right){\bf{v}}_j^H } \right\|^2 }}}  + K\gamma }}.
\end{align}
\fi
In order to simplify the denominator in (\ref{f53}), we approximate it as
\begin{equation}
\label{f54}
\left\| {\left( {{\bf{I}} - {\bf{\tilde H}}_j^H {\bf{\tilde H}}_j } \right){\bf{v}}_j^H } \right\|^2  \approx \left\| {\left( {{\bf{I}} - {\bf{\tilde H}}_i^H {\bf{\tilde H}}_i } \right){\bf{v}}_i^H } \right\|,1 \leq i,j \leq K.
\end{equation}
When $K=2$, the left side is strictly equal to the right side in (\ref{f54}). When $K > 2$, as ${\bf{v}}_i ,i = 1, \cdots ,K$, are independent and uniformly distributed in ${\mathbb G}\left( {1,n} \right)$, the left side and right side of (\ref{f54}) approach to the same value as $K$ grows. It means that (\ref{f54}) is a reasonable approximation for a large value of $K$. With (\ref{f54}), (\ref{f53}) could be further reduced to (\ref{f55}) as
\begin{equation}
\label{f55}
{\mathrm{SINR}}_i^{{\text{ZF,non - ideal}}}  = \frac{{\left\| {{\bf{u}}_i \left( {{\bf{I}} - {\bf{\tilde H}}_i^H {\bf{\tilde H}}_i } \right){\bf{v}}_i^H } \right\|^2 }}
{{\sum\limits_{\scriptstyle j = 1 \hfill \atop 
  \scriptstyle j \ne i \hfill} ^K {\left\| {{\bf{u}}_i \left( {{\bf{I}} - {\bf{\tilde H}}_j^H {\bf{\tilde H}}_j } \right){\bf{v}}_j^H } \right\|^2 }  + \gamma _{eff} }}
\end{equation}
where $\gamma _{eff}  = K\left\| {\left( {{\bf{I}} - {\bf{\tilde H}}_i^H {\bf{\tilde H}}_i } \right){\bf{v}}_i^H } \right\|^2 \gamma $. Substituting (\ref{f42}) into (\ref{f55}), \ifCLASSOPTIONonecolumn
we get
\begin{equation}
\label{f56}
{\mathrm{SINR}}_i^{{\mathrm{MU,non - ideal}}}  = \frac{{\left\| {\alpha e^{j\theta _i }  - \alpha e^{j\theta _i } {\bf{v}}_i {\bf{\tilde H}}_i^H {\bf{\tilde H}}_i {\bf{v}}_i^H  + \sqrt {1 - \alpha ^2 } e^{jw_i } {\bf{v}}_i^ \bot  {\bf{\tilde H}}_i^H {\bf{\tilde H}}_i {\bf{v}}_i^H } \right\|^2 }}
{{\sum\limits_{\scriptstyle j = 1 \hfill \atop 
  \scriptstyle j \ne i \hfill} ^K {\left\| {\sqrt {1 - \alpha ^2 } e^{jw_i } {\bf{v}}_i^ \bot  {\bf{v}}_j^H  - \sqrt {1 - \alpha ^2 } e^{jw_i } {\bf{v}}_i^ \bot  {\bf{\tilde H}}_{i,j}^H {\bf{\tilde H}}_{i,j} {\bf{v}}_j^H } \right\|^2 }  + \gamma _{eff} }}.
\end{equation}
\else
we get (\ref{f56}) at the top of the next page.
\begin{figure*}[t]
\normalsize
\setcounter{MYtempeqncnt}{\value{equation}}
\setcounter{equation}{55}
\hrulefill
\vspace*{4pt}
\begin{equation}
\label{f56}
{\mathrm{SINR}}_i^{\mathrm{MU,non-ideal}} = \frac{{{{\left\| {\alpha {e^{j{\theta _i}}} - \alpha {e^{j{\theta _i}}}{\bf{v}}_i^{}{\bf{\hat H}}_i^H{\bf{\hat H}}_i^{}{\bf{v}}_i^H + \sqrt {1 - {\alpha ^2}} {e^{j{w_i}}}{\bf{v}}_i^ \bot {\bf{\hat H}}_i^H{\bf{\hat H}}_i^{}{\bf{v}}_i^H} \right\|}^2}}}{{\sum\limits_{\scriptstyle j = 1 \hfill \atop 
  \scriptstyle j \ne i \hfill} ^K {{{\left\| {\left( {\sqrt {1 - {\alpha ^2}} {e^{j{w_i}}}{\bf{v}}_i^ \bot {\bf{v}}_j^H - \sqrt {1 - {\alpha ^2}} {e^{j{w_i}}}{\bf{v}}_i^ \bot {\bf{\hat H}}_{i,j}^H{\bf{\hat H}}_{i,j}^{}{\bf{v}}_j^H} \right)} \right\|}^2}}  + {\gamma _{eff}} }}
\end{equation}
\setcounter{equation}{\value{MYtempeqncnt}}
\addtocounter{equation}{1} 
\end{figure*}
\fi
Since the above formula is still too complicated, we continue to simplify the numerator and denominator respectively by ignoring the secondary factors. For the numerator, as the probability of ${\bf{v}}_i^ \bot  {\bf{\tilde H}}_i^H {\bf{\tilde H}}_i {\bf{v}}_i^H  \ll {\bf{v}}_i {\bf{\tilde H}}_i^H {\bf{\tilde H}}_i {\bf{v}}_i^H$ approaches to 1 when $K \geq 2$, the term $\sqrt {1 - \alpha ^2 } e^{jw_i } {\bf{v}}_i^ \bot  {\bf{\tilde H}}_i^H {\bf{\tilde H}}_i {\bf{v}}_i^H$ can be dropped and the result becomes $\alpha ^2 \left( {1 - \sum\limits_{\scriptstyle j = 1 \hfill \atop \scriptstyle j \ne i \hfill} ^K {\left| {{\bf{v}}_i {\bf{v}}_j^H } \right|^2 } } \right)^2 $. Similarly for the denominator, we reduce it to $\left( {1 - \alpha ^2 } \right)\sum\limits_{\scriptstyle j = 1 \hfill \atop 
   \scriptstyle j \ne i \hfill} ^K {\left| {{\bf{v}}_i^ \bot  {\bf{v}}_j^H } \right|^2 }  + K\left( {1 - \sum\limits_{\scriptstyle k = 1 \hfill \atop 
   \scriptstyle k \ne j \hfill} ^K {\left| {{\bf{v}}_k {\bf{v}}_j^H } \right|^2 } } \right)\gamma$. In addition, we have $\left| {{\bf{v}}_i^ \bot  {\bf{v}}_j^H } \right| \approx \left| {{\bf{v}}_i {\bf{v}}_j^H } \right|$. Therefore, the SINR of $i$th user could be rewritten as
   \begin{equation}
   \label{f57}
   {\mathrm{SINR}}_i^{{\mathrm{ZF,non - ideal}}}  = \frac{{\alpha ^2 \left( {1 - \sum\limits_{\scriptstyle j = 1 \hfill \atop 
     \scriptstyle j \ne i \hfill} ^K {\left| {{\bf{v}}_i {\bf{v}}_j^H } \right|^2 } } \right)^2 }}
   {{\left( {1 - \alpha ^2  - K\gamma } \right)\sum\limits_{\scriptstyle j = 1 \hfill \atop 
     \scriptstyle j \ne i \hfill} ^K {\left| {{\bf{v}}_i {\bf{v}}_j^H } \right|^2 }  + K\gamma }}.
   \end{equation}
With (\ref{f57}), the BS can choose the suitable combination of modulation scheme and channel coding rate for each user in a MU-MIMO user group. As a result, the BS can properly group MU-MIMO users and adaptively switch between MU-MIMO and SU-MIMO.

In some cases, the expectation of SINR provides a more reliable prediction of the MU-MIMO spectral efficiency at a given error level of CSI. Let $z = \sum\limits_{\scriptstyle j = 1 \hfill \atop \scriptstyle j \ne i \hfill} ^K {\left| {{\bf{v}}_i {\bf{v}}_j^H } \right|^2 }  = \sum\limits_{\scriptstyle j = 1 \hfill \atop \scriptstyle j \ne i \hfill} ^K {x_j^2 } $, then ${\mathrm{SINR}}^{{\text{ZF,non - ideal}}} $ could be written as a function of $z$: ${\mathrm{SINR}}^{{\mathrm{ZF,non - ideal}}}  = g\left( z \right)$. Since $g\left( z \right)$ has a form of
 $g\left( z \right) = az + b + \frac{c}{z}$, it is a convex function. Thus by making using of Jensen inequality (${\mathbb E}\left( {g\left( x \right)} \right) \geq g\left( {{\mathbb E}\left( x \right)} \right)$ when $g\left( x \right)$ is a convex function), the expectation of (\ref{f57}) has a lower bound
\begin{equation}
\label{f58}
{{\mathbb E}}\left[ {{\mathrm{SINR}}_i^{{\mathrm{ZF,non - ideal}}} } \right] \geq \frac{{\alpha ^2 \left( {1 - {\mathbb E}\left[ z \right]} \right)^2 }}
{{\left( {1 - \alpha ^2  - K\gamma } \right){\mathbb E}\left[ z \right] + K\gamma }}.
\end{equation}
As $x_j ,j = 1, \cdots ,K,j \ne i$, are i.i.d. random variables, according to (\ref{f48}), we have ${\mathbb E}\left[ z \right] = (K - 1)(N - 1) B (2,N - 1)$, and we obtain Theorem 2 as summarized below.

\begin{theorem}
For an $\left( {N,K,\alpha } \right)$ MU-MIMO communication system, where $N$ is the number of transmit antenna at the BS side, and $K$ is the number of grouped users. If the ${\mathrm{SINR}}^{{\mathrm{SU}}}$ of the $i$th user is ${1 \mathord{\left/
 {\vphantom {1 \gamma }} \right.\kern-\nulldelimiterspace} \gamma }$ which is defined as ${{P\left\| {{\bf{h}}_i } \right\|^2 } \mathord{\left/{\vphantom {{P\left\| {{\bf{h}}_i } \right\|^2 } {\sigma _{NI}^2 }}} \right.\kern-\nulldelimiterspace} {\sigma _{NI}^2 }}$, then the expected SINR at the receiver has a lower bound
\ifCLASSOPTIONonecolumn
\begin{equation}
\label{f59}
{\mathbb E}\left[ {{\mathrm{SINR}}_i^{{\mathrm{MU,non - ideal}}} \left( \alpha  \right)} \right] \geq \frac{{\alpha ^2 \left( {N - K + 1} \right)}}
{{\left( {1 - \alpha ^2  - K\gamma } \right)\left( {K - 1} \right) + NK\gamma }}.
\end{equation}
\else 
\begin{align}
\label{f59}
& {\mathbb E}\left[ {{\mathrm{SINR}}_i^{{\mathrm{MU,non - ideal}}} \left( \alpha  \right)} \right] \nonumber \\ 
& \geq \frac{{\alpha ^2 \left( {N - K + 1} \right)}}
{{\left( {1 - \alpha ^2  - K\gamma } \right)\left( {K - 1} \right) + NK\gamma }}.
\end{align} 
\fi
\end{theorem}
With (\ref{f59}), we could estimate the expectation of channel capacity of MU-MIMO when ZF precoding is used for large-scale MU-MIMO systems. 

\subsubsection{Discussion}
at this point, we provide some discussion about the approximations (\ref{f57}) and (\ref{f59}). For (\ref{f57}), there are three factors affecting its accuracy.

Firstly, given the values of $K$ and $\alpha$, the probability of ${\bf{\tilde H}}^{\bf{H}} {\bf{\tilde H}}_k  \to {\bf{I}}$ increases as $N$ grows in (\ref{f51}), hence the approximation error of the numerator in (\ref{f56}) is almost ignorable when $N$ is large, e.g., $N=100$. On the other hand, the values of ${\bf{v}}_i^ \bot  {\bf{\tilde H}}_i^H {\bf{\tilde H}}_i {\bf{v}}_i^H $ and ${\bf{v}}_i^ \bot  {\bf{\tilde H}}_i^H {\bf{\tilde H}}_{i,j} {\bf{v}}_i^H $ are closer and closer to zero as $N$ grows. As a result, the approximations of the numerator and denominator in (\ref{f56}) are reasonable for large $N$. Therefore, we conclude that the gap between the approximated value in (\ref{f57}) and the real SINR becomes smaller as $N$ grows. 

Secondly, given the values of $N$ and $\alpha$, the probability of ${\bf{\tilde H}}^{\bf{H}} {\bf{\tilde H}}_k  \to {\bf{I}}$ decreases as $K$ grows, while the values of ${\bf{v}}_i^ \bot  {\bf{\tilde H}}_i^H {\bf{\tilde H}}_i {\bf{v}}_i^H $ and ${\bf{v}}_i^ \bot  {\bf{\tilde H}}_i^H {\bf{\tilde H}}_{i,j} {\bf{v}}_i^H $ show inverse tendency. This means that the approximation of ${\bf{\tilde H}}^{\bf{H}} {\bf{\tilde H}}_k  = {\bf{I}}$ in (\ref{f51}) and the omission of ${\bf{v}}_i^ \bot  {\bf{\tilde H}}_i^H {\bf{\tilde H}}_i {\bf{v}}_i^H $ and ${\bf{v}}_i^ \bot  {\bf{\tilde H}}_i^H {\bf{\tilde H}}_{i,j} {\bf{v}}_i^H $ in (\ref{f56}) would cause larger error as $K$ grows. 

Thirdly, given the values of $N$ and $K$, the value of $\left| {\sqrt {1 - \alpha ^2 } e^{jw_i } {\bf{v}}_i^ \bot  {\bf{\tilde H}}_i^H {\bf{\tilde H}}_i {\bf{v}}_i^H } \right|$ decreases as $\alpha$ increases. Hence, ignoring this term causes little error in (\ref{f56}) for $\alpha$ close to 1. Therefore, as the CSI is more accurate, the approximation in (\ref{f57}) resluts in less error.

In conclusion, larger values of $N$, $\alpha$, or smaller value of $K$ result in a more accurate approximation in (\ref{f57}).

As (\ref{f59}) provides a lower bound of expectation of SINR, we can analyze the factor which affects the gap between the lower bound and the real value. Firstly, the term $\sum\limits_{j = 1}^{K - 1} {x_j^2 } $ approaches to 
$\left( {K - 1} \right){\mathbb E}\left[ {x_j^2 } \right]$ as $K$ grows, which means that for a large value of $K$, the inequality in (\ref{f59}) approaches to equality. Secondly, if $\sum\limits_{j = 1}^{K - 1} {x_j^2 } $ is negligible relative to the value of $K\gamma $, e.g., $N$ or $\gamma$ is very large, the inequality also approaches to equality. As a result, the lower bound of (\ref{f59}) is asymptotically tight as $K$ or $N$ grows.

With the approximation of SINR in (\ref{f57}) and ${\mathbb E}\left[ {{\text{SINR}}} \right]$ in (\ref{f59}), it is easy for the BS to predict the capacity of each user in MU-MIMO mode, the sum capacity gain over SU-MIMO, and the total capacity of the grouped users.

\subsubsection{Numerical Results}
\begin{figure}[!t]
\centering \includegraphics[width = 0.8\linewidth]{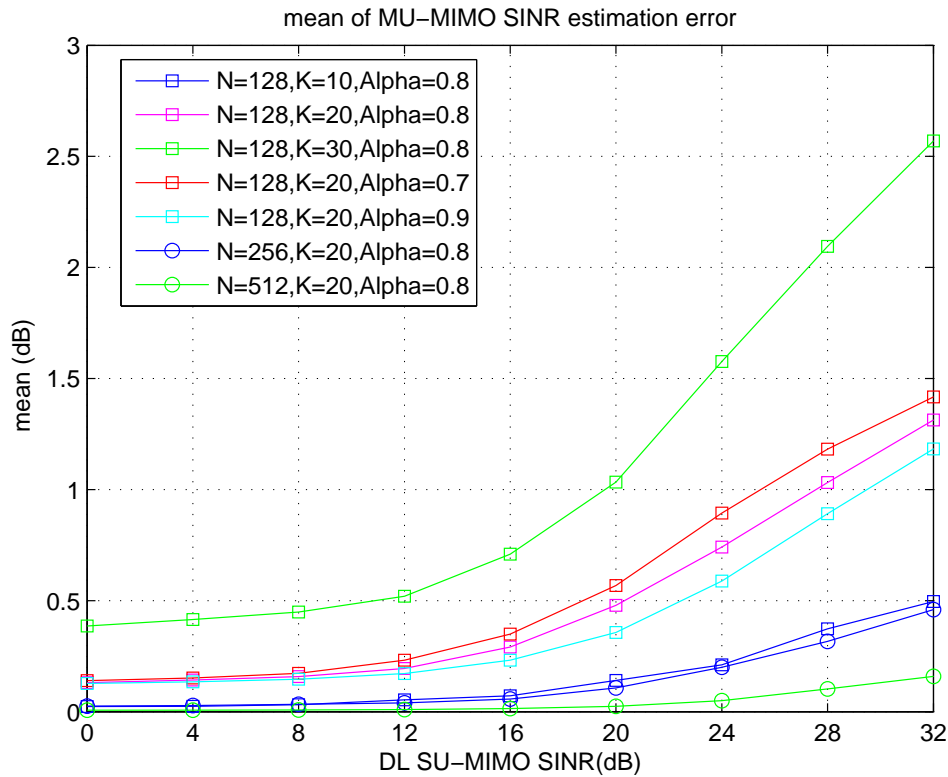}
\caption{Mean of the estimation error of (\ref{f57}) }
\end{figure}

\begin{figure}[!t]
\centering \includegraphics[width = 0.8\linewidth]{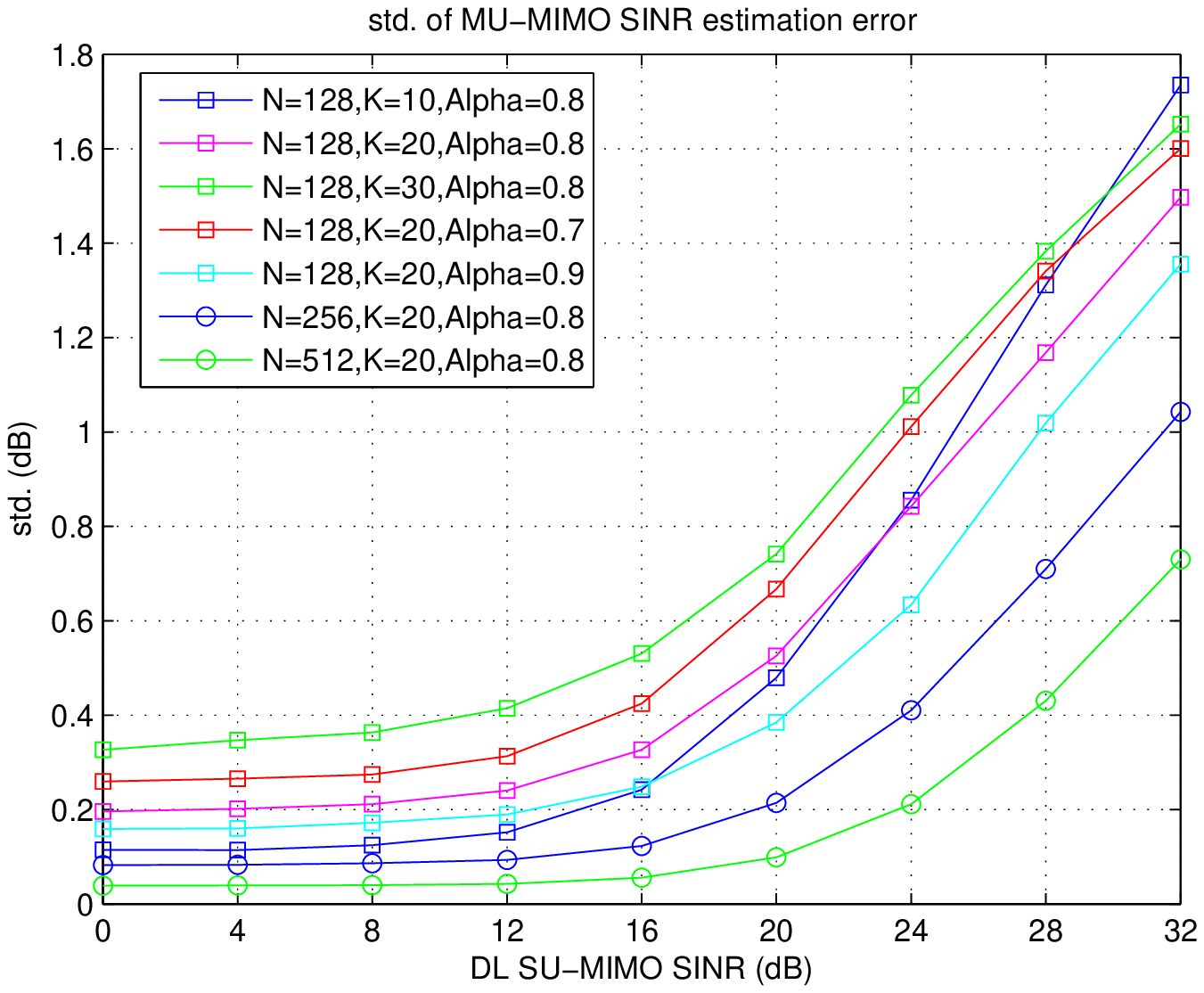}
\caption{Std. of the estimation error of (\ref{f57}) }
\end{figure}

\begin{figure}[!t]
\centering \includegraphics[width = 0.8\linewidth]{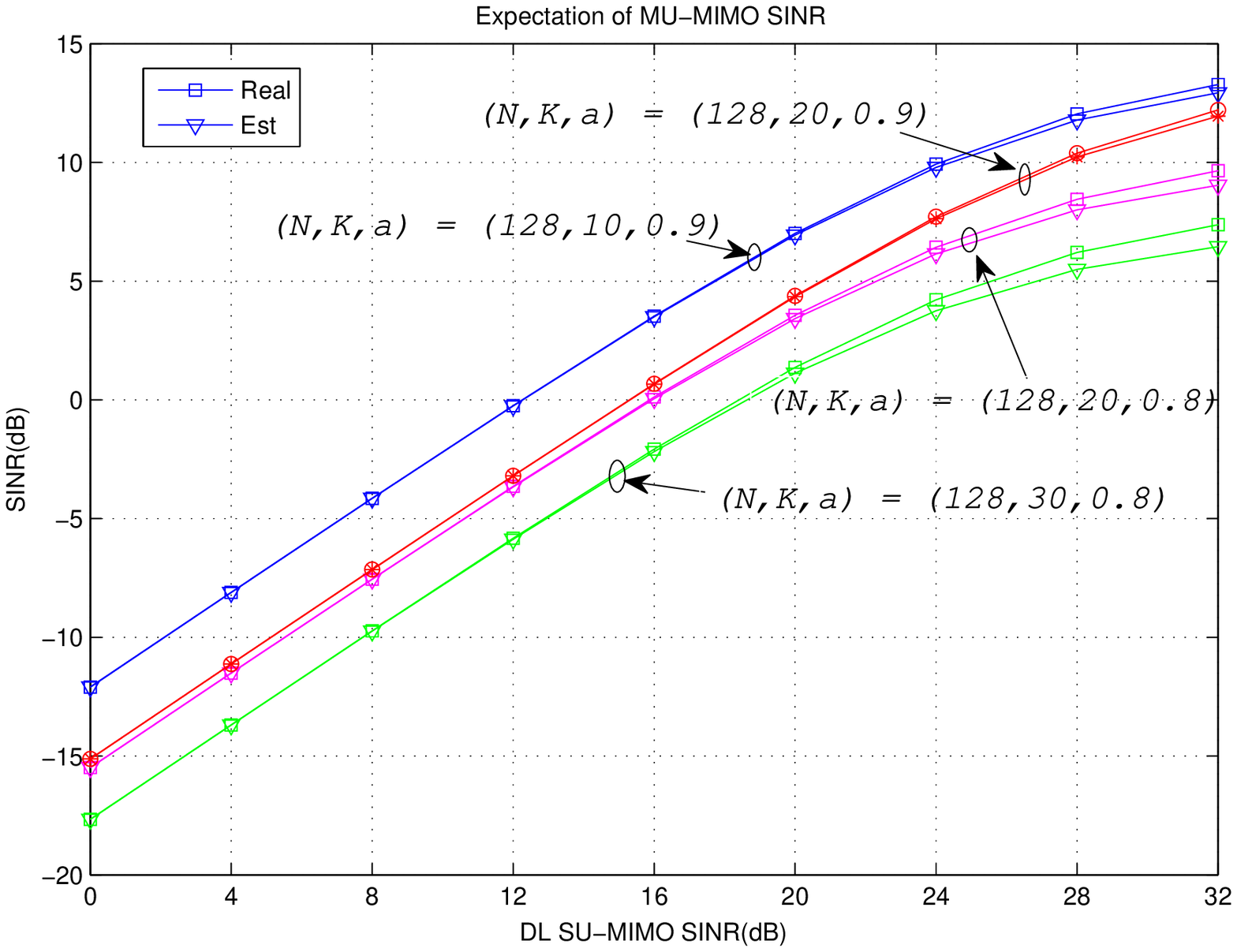}
\caption{Comparison between (\ref{f59}) and expectation of SINR}
\end{figure}

we present numerical simulation results in this section to verify the validity of (\ref{f57}) and (\ref{f59}). Similarly to the conjugate beamforming case, the mean and std. of the estimation error are used to measure the accuracy of (\ref{f57}), and the results with different values of $K$, $N$, and $\alpha$ are shown in Fig. 8 and Fig. 9 respectively. For (\ref{f59}), the estimation value and the real expectation of SINR are compared in Fig. 10.  

In Fig. 8, we can see that the mean of estimation error increases as the SU-MIMO SINR increases, which is because when the value of SU-MIMO SINR is small, the noise and inter-cell interference are the major factors affecting the MU-MIMO SINR. As the value of SU-MIMO SINR increases, the residual multi-user interference becomes the major factor, and the estimation error of (\ref{f57}) becomes dominant. In addition, the mean of estimation error increases as the value of $K$ increases, and decrease as the value of $N$ or $\alpha$ decreases, which verifies the analysis in 2) above. For the same reasons, Fig. 9 shows similar variation tendency to Fig. 8. Considering that the number of grouped users is about 1/10 to 1/5 of the number of antennas in large-scale MU-MIMO systems (e.g., $K=20$, $N=128$), (\ref{f57}) provides SINR estimation with relatively high accuracy.

In Fig. 10, the difference between the estimation and real values is close to 0 for most cases, and it is about 1 dB in the high SINR region. Hence, it provides a reasonable estimation of the expectation of SINR.

\subsection{CB vs. ZF}

\subsubsection{SINR Gain of ZF over CB}
in this section, we compare the SINRs of CB and ZF in (44) and (59). The SINR gain of ZF over CB at a given CSI error level is
\begin{equation}
\label{f60}
G^{{\mathrm{ZF - CB}}}  = \frac{{{\mathrm{SINR}}^{{\mathrm{ZF,non-ideal}}} }}{{{\mathrm{SINR}}^{{\text{CB,non - ideal}}} }} = \frac{{\left( {1 - z} \right)^2 \left( {z + K\gamma } \right)}}{{\left( {1 - \alpha ^2  - K\gamma } \right)z + K\gamma }}
\end{equation}
where $z$ is the same as in (\ref{f58}) which denotes the multi-user interference power among the users in MU-MIMO. According to (\ref{f60}), $z \to \left( {K - 1} \right)\left( {N - 1} \right)B\left( {2,N - 1} \right)$ when $K$ is large, then (\ref{f60}) is reduced to a simple function of $K$, $N$, and $\alpha$ as
 
\begin{equation}
\label{f61}
G^{{\mathrm{ZF - CB}}}  \to \frac{1}{{N^2 }} \cdot \frac{{\left( {N - K + 1} \right)\left( {K + NK\gamma  - 1} \right)}}{{\left( {1 - \alpha ^2  - K\gamma } \right)K\gamma  + NK\gamma }}.
\end{equation}

Here, we offer a discussion about (\ref{f60}) and (\ref{f61}). In the high ${\mathrm{SINR}}^{{\mathrm{SU}}}$ region, as $\gamma  \to 0$ and $\gamma  \ll z$, we have $ G^{{\mathrm{ZF - CB}}}  \to \frac{{\left( {1 - z} \right)^2 }}{{1 - \alpha ^2 }}$, and ${\mathbb{E}}\left[ {{G^{{\mathrm{ZF - CB}}}}} \right] \to \frac{{{N^2} + \left( {K - 1} \right)\left( {K - 1 - 2N} \right)}}{{{N^2}\left( {1 - {\alpha ^2}} \right)}}$. Therefore, the SINR gain of ZF over CB increases from $\left( {1 - z} \right)^2 $ to positive infinity when $\alpha$ varies from 0 to 1. It indicates that ZF has an overwhelming advantage compared to CB when the CSI is ideal, which coincides with our conventional understanding. However, as the error of CSI increases, the SINR gain of ZF begins to decrease, and both ZF and CB has the same SINR value when $\alpha  = \sqrt {2z - z^2 } $. If $\alpha$ decreases further, the SINR of ZF becomes smaller than that of CB. 

In the low ${\mathrm{SINR}}^{{\mathrm{SU}}}$ region, as $\gamma  \gg z$, $G^{{\mathrm{ZF - CB}}}  \to 1 - z$, and ${\mathbb{E}}\left[ {{G^{{\mathrm{ZF - CB}}}}} \right] \to \frac{{N - K + 1}}{N}$. It means that the SINR gain of ZF over CB is definitely smaller than 1, regardless of the considered parameters.

Given the value of $\alpha$, increasing the value of $N$ or decreasing the value of $K$ could result in smaller $z$ according to the definition of $z$, thus enlarge the SINR gain of ZF over CB.

In conclusion, if the CSI is not ideal, the SINR gain of ZF over CB is determined by the parameter $\alpha$ given the values of $N$ and $K$ in the high ${\mathrm{SINR}}^{{\mathrm{SU}}}$ region, while the SINR gain of ZF is always smaller than 1 in the low ${\mathrm{SINR}}^{{\mathrm{SU}}}$ region and independent of $\alpha$.

\subsubsection{Numerical Results}
\begin{figure}[!t]
\centering \includegraphics[width = 0.8\linewidth]{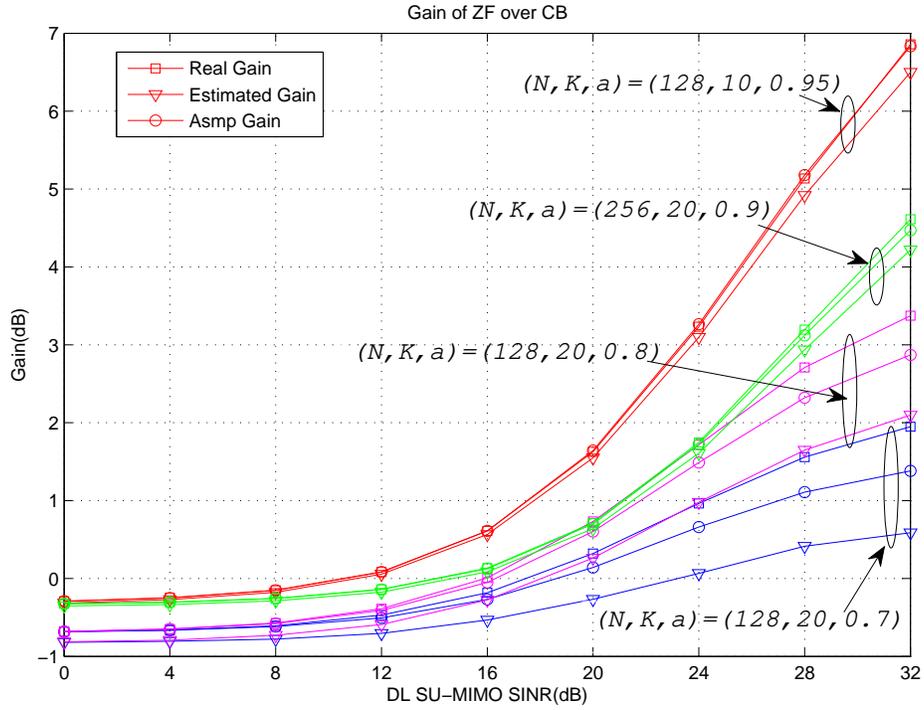}
\caption{Gain of ZF over CB}
\end{figure}

\begin{figure}[!t]
\centering \includegraphics[width = 0.8\linewidth]{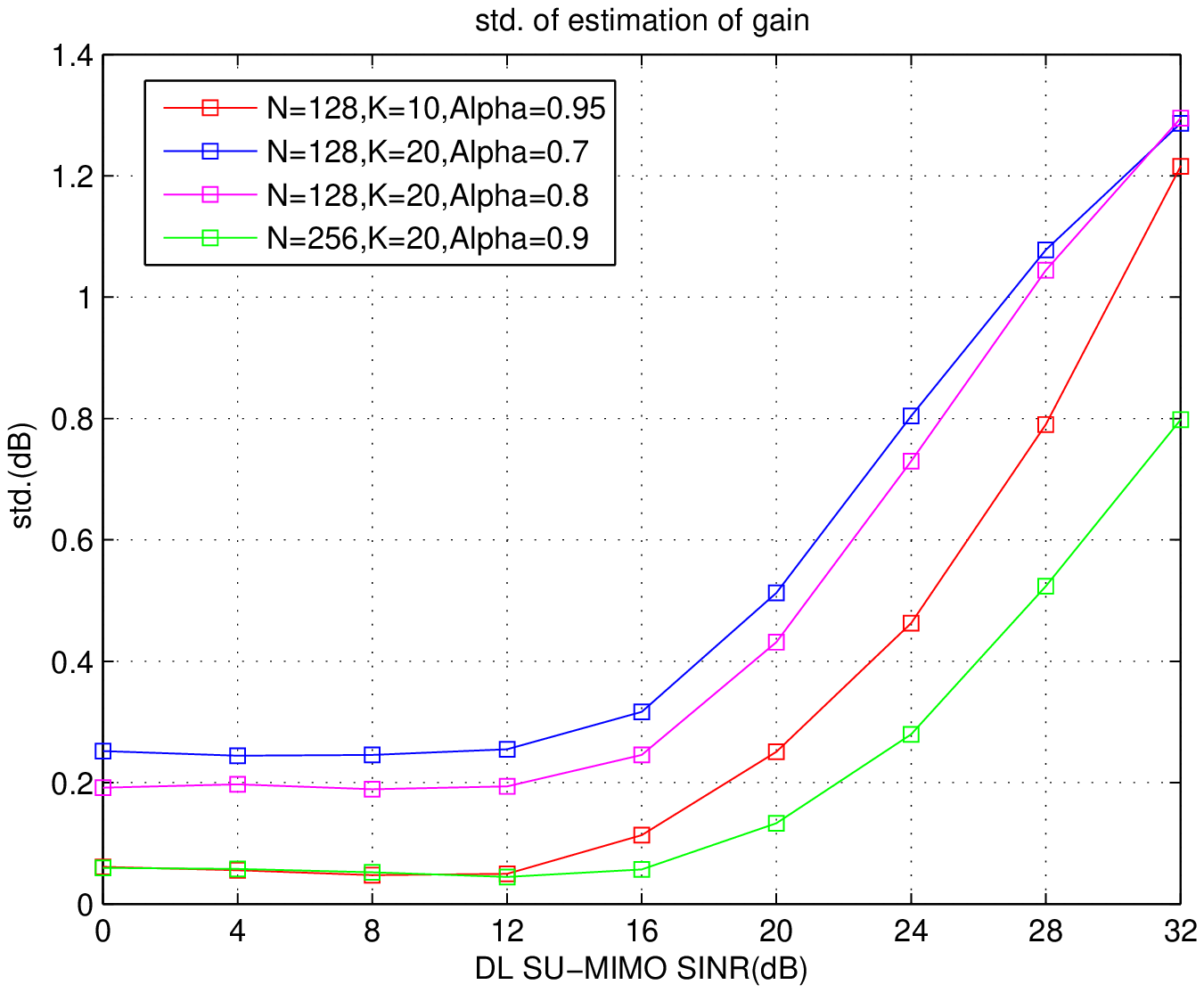}
\caption{Std. of estimation, formula (\ref{f60})}
\end{figure}

\begin{figure}[!t]
\centering \includegraphics[width = 0.8\linewidth]{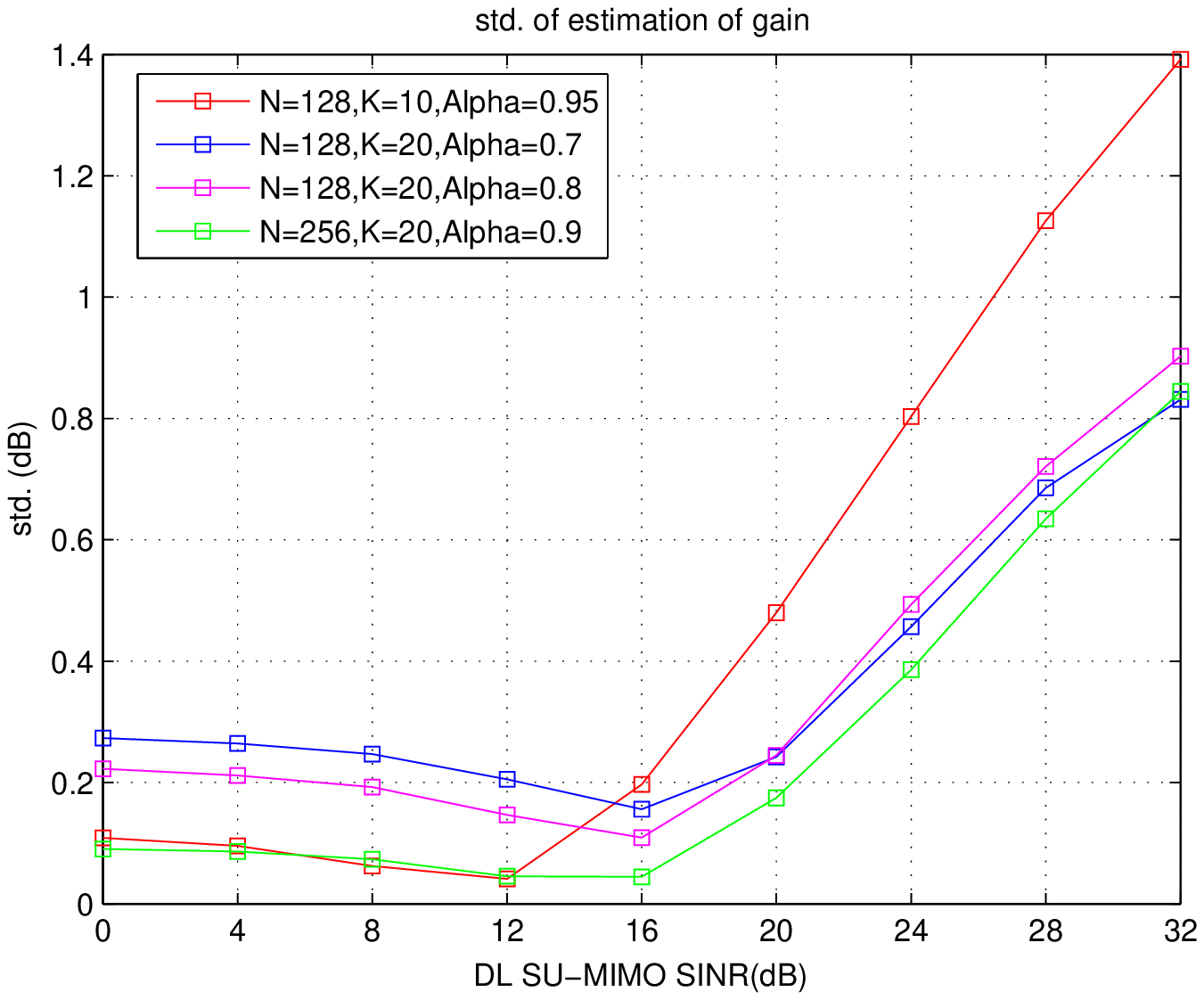}
\caption{Std. of estimation, formula (\ref{f61}) }
\end{figure}

we provide some numerical results to verify the rationality of formula (\ref{f60}) and (\ref{f61}) in Fig. 11, Fig. 12, and Fig. 13. In Fig. 11, the formulas (\ref{f60}) (Estimated Gain) and (\ref{f61}) (Asmp Gain) are compared with the real gain, where we can see that both the formulas (\ref{f60}) and (\ref{f61}) provide underestimation of the real gain. Nevertheless, the estimation errors are relatively small, and these two formulas could be used to predict the gain of ZF over CB for large-scale MU-MIMO systems. The std. of (\ref{f60}) and (\ref{f61}) are shown in Fig. 12 and Fig. 13 respectively. For the same reason, the variation of std. with parameters $K$, $N$, and $\alpha$ in Fig. 12 is similar to that of Fig. 9. Fig. 13 shows a relatively different variation of std., where the std. increases rapidly as $K$ decreases because (\ref{f61}) can be considered accurate only when $K$ is large. In summery, these figures indicate that formulas (\ref{f60}) and (\ref{f61}) can be used to predict the gain of ZF over CB, and switch between these two precoding methods adaptively.

\subsection{ZF with Ideal CSI}

\subsubsection{Estimation of SINR}
when the CSI is ideal, the multi-user interference is removed completely when ZF precoding is employed, then the SINR of the $i$th user for MU-MIMO is 
\begin{equation}
\label{f62}
{\mathrm{SINR}}_i^{{\text{ZF,ideal}}}  = \frac{{\left\| {{\bf{h}}_i {\bf{w}}_i } \right\|^2 }}{{\sigma _{NI}^2 }}
\end{equation}
where ${\bf{w}}_i  = \frac{{\left( {{\mathbf{I}} - {\mathbf{\tilde H}}_i^H {\mathbf{\tilde H}}_i } \right){\mathbf{u}}_i^H }}{{\left\| {\left( {{\bf{I}} - {\mathbf {\tilde H}}_i^H {\mathbf{\tilde H}}_i } \right){\mathbf{u}}_i^H } \right\|}}$ with the matrix ${\mathbf{\tilde H}}_i$ defined as ${\mathbf{\tilde H}}_i  = [{\mathbf{u}}_1^T  \, \cdots \, {\mathbf{u}}_{i - 1}^T  \, {\mathbf{u}}_{i + 1}^T  \, \cdots \, {\mathbf{u}}_K^T ]^T $. Note that the factor
 $\left( {{\mathbf{\tilde H}}_i {\mathbf{\tilde H}}_i^H } \right)^{ - 1}$ is ignored in ${\bf{w}}_i$ here for the same reason as in (\ref{f52}). Substituting ${\bf{w}}_i$ into (\ref{f62}) and after simplification, we have 
 \begin{equation}
 \label{f63}
 {\mathrm{SINR_i}}^{{\text{ZF,ideal}}}  = \frac{{1 - {\bf{u}}_i {\mathbf{\tilde H}}_i {\mathbf{\tilde H}}_i^H {\bf{u}}_i }}
 {{K\gamma }}.
 \end{equation}
Using the same method as in Section IV.A, the expectation of ${\text{SINR}}^{{\text{MU,ideal}}} $ is calculated as
\ifCLASSOPTIONonecolumn
\begin{equation}
\label{f64}
{\mathbb E}\left[ {{\mathrm{SINR}}_i^{{\text{ZF,ideal}}} } \right] = \frac{{1 - {\mathbb E}\left[ {{\bf{u}}_i {\mathbf{\tilde H}}_i {\mathbf{\tilde H}}_i^H {\bf{u}}_i } \right]}}
{{K\gamma }} = \frac{{1 - \left( {K - 1} \right){\mathbb E}\left[ {x^2 } \right]}}
{{K\gamma }} = \frac{{N - K + 1}}
{{NK\gamma }}.
\end{equation}
\else
\begin{align}
\label{f64}
{\mathbb E}\left[ {{\mathrm{SINR}}_i^{{\text{ZF,ideal}}} } \right] & = \frac{{1 - {\mathbb E}\left[ {{\bf{u}}_i {\mathbf{\tilde H}}_i {\mathbf{\tilde H}}_i^H {\bf{u}}_i } \right]}}
{{K\gamma }} \nonumber \\ & = \frac{{1 - \left( {K - 1} \right){\mathbb E}\left[ {x^2 } \right]}}
{{K\gamma }} \nonumber \\ & = \frac{{N - K + 1}}
{{NK\gamma }}.
\end{align}
\fi
It is not difficult to find that the value of (\ref{f64}) is equal to that of (\ref{f59}) when $\alpha=1$. 

\subsubsection{Numerical Results}
Fig. 14 and Fig. 15 show the accuracy of formula (\ref{f63}), where both the mean and std. of estimation error is very close to zero. Hence, formula (\ref{f63}) provides a very accurate upper bound of the SINR of MU-MIMO given $
{\mathrm{SINR}}^{{\mathrm{SU}}}$. In Fig. 16, the estimation matches the real value almost perfectly. Therefore, formula (\ref{f64}) could be used to estimate ergodic capacity given ${\mathrm{SINR}}^{{\mathrm{SU}}} $.

\begin{figure}[!t]
\centering \includegraphics[width = 0.8\linewidth]{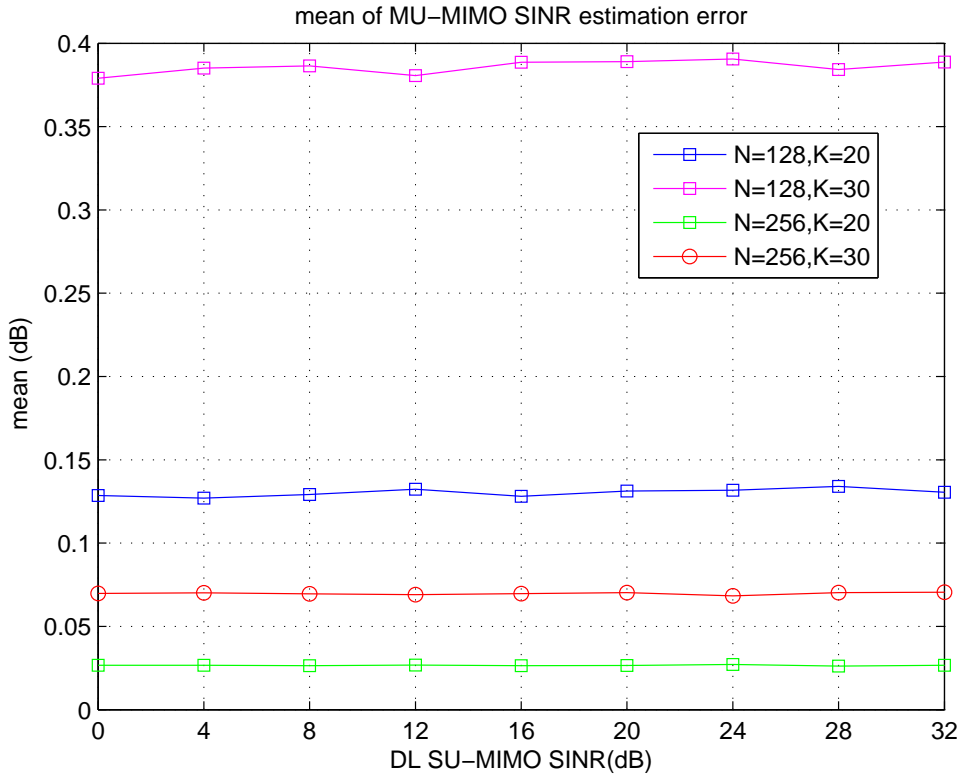}
\caption{Mean of estimation error, formula (\ref{f63})}
\end{figure}

\begin{figure}[!t]
\centering \includegraphics[width = 0.8\linewidth]{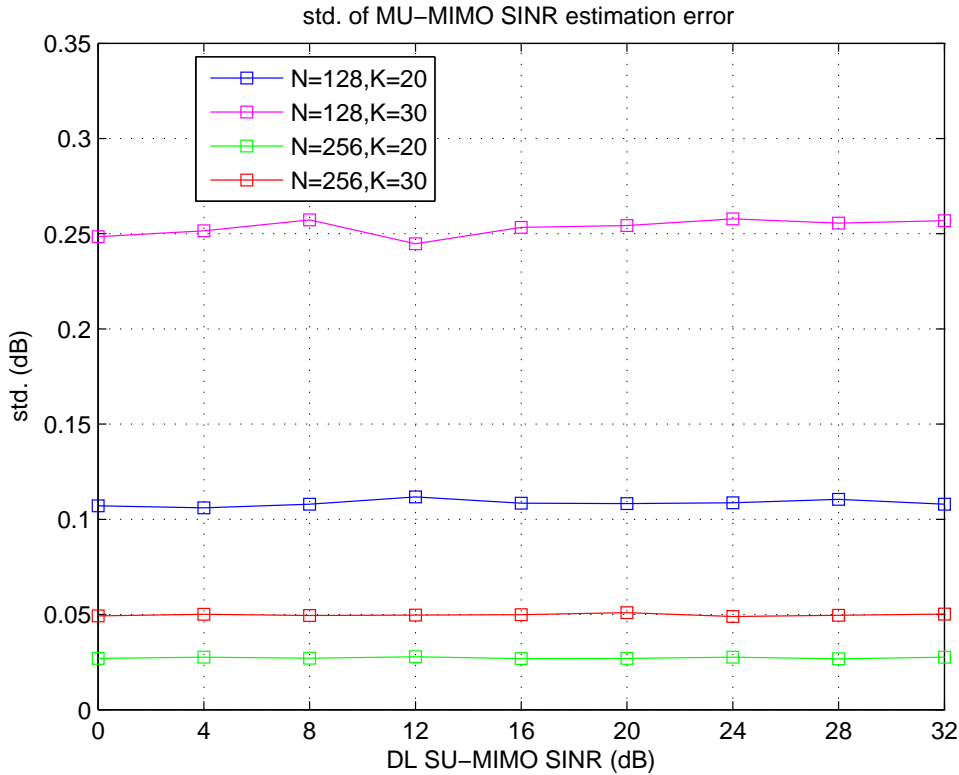}
 \caption{Std. of estimation error, formula (\ref{f63})}
\end{figure}

\begin{figure}[!t]
\centering \includegraphics[width = 0.8\linewidth]{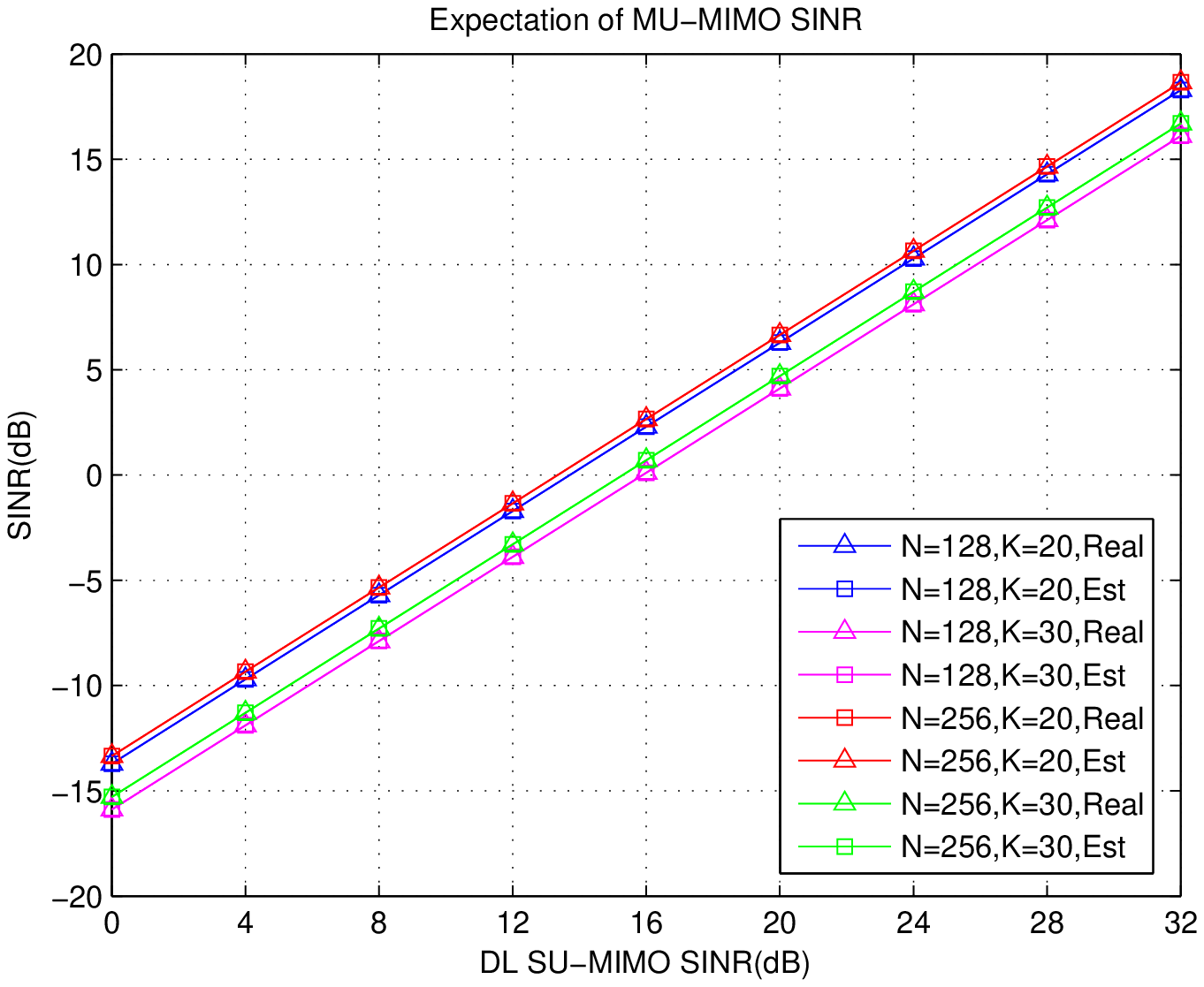}
\caption{Expectation of SINR, formula (\ref{f64})}
\end{figure}

\section{Conclusion}
We studied the volume of a hyperball in a complex Grassmann manifold based on the probability of canonical angles between any two points and obtained closed-form formulas for various $\left( {k,n} \right)$ values and radii of the hyperball. In addition, we provided an application of the formulas in large-scale MU-MIMO communication systems and derived closed-form approximations of MU-MIMO SINRs given the number of BS antennas $N$, MU-MIMO grouped user number $K$, CSI error $\alpha$, and the user channel quality ${\mathrm{SINR}}^{{\mathrm{SU}}}$ in SU-MIMO mode. Simulation results verified the accuracy of the formulas and the closed-form approximations. Our results solve a fundamental problem whose solution has been missing but is necessary for a practical deployment of massive MU-MIMO systems. In the future, the cases of more than one receive antennas and correlated channel matrix will be considered, where the volume formulas of hyperball in Grassmann manifold when $k \geq 2$ will be applied.

\appendices

\section{Proof Of Theorem 1}
\begin{IEEEproof}
\begin{figure}[!t]
\centering \includegraphics[width = 0.8\linewidth]{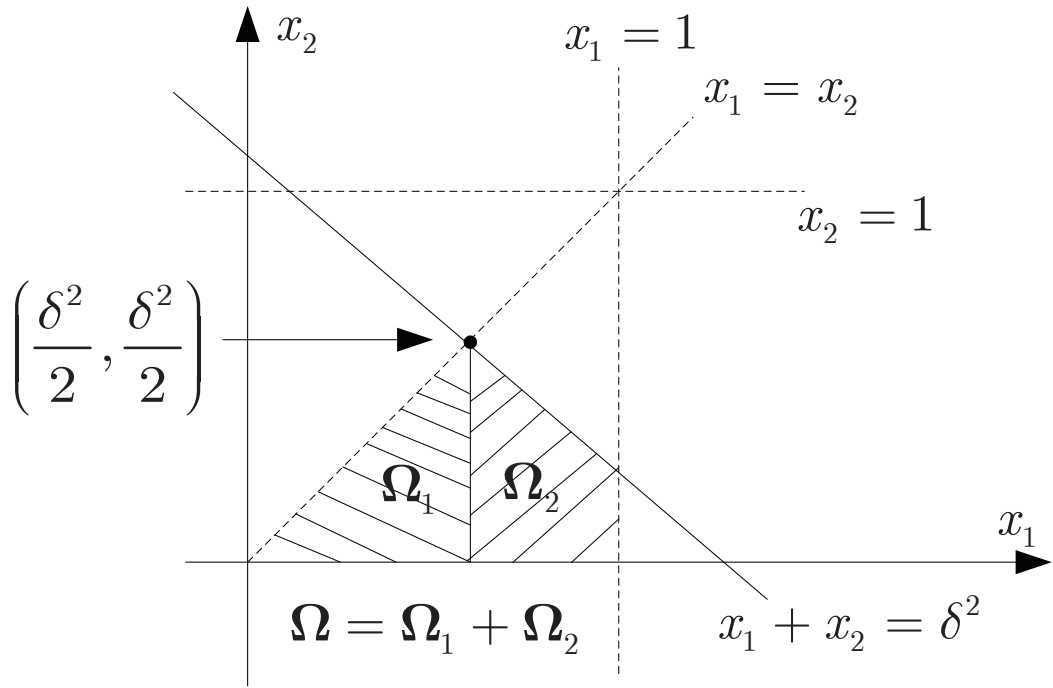}
\caption{{The domain of integration in (\ref{f66})}}
\end{figure}

Let ${\sin ^2}{\theta _1} = {x_1}$ and ${\sin ^2}{\theta _2} = {x_2}$, then the probability could be written as
\begin{equation}
\label{f65}
p\left( {{d_{pF}} \leq \delta } \right) = M\iint\limits_{\mathbf{\Omega }} {x_1^{n - 2k}x_2^{n - 2k}{{\left( {{x_1} - {x_2}} \right)}^2}d{x_1}d{x_2}}.
\end{equation}
When $0 \leq {\delta ^2} \leq 1$, the domain of integration is the set $ {\mathbf{\Omega }} = \left\{ {{x_1} + {x_2} < \delta^2 ,0 < {x_1},{x_2} < 1} \right\}$
which is shown in Fig. 17, where ${\mathbf{\Omega }}$ could be divided into two parts ${\mathbf{\Omega_1 }}$ and ${\mathbf{\Omega_2 }}$. The sum integration on these two parts is
\ifCLASSOPTIONonecolumn
\begin{equation}
\label{f66}
\begin{split}
&\iint\limits_{{{\bf{\Omega }}_1} + {{\bf{\Omega }}_2}} {x_1^{n - 4}x_2^{n - 4}{{\left( {{x_1} - {x_2}} \right)}^2}d{x_1}d{x_2}}\\
  & = \iint\limits_{{{\bf{\Omega }}_1} + {{\bf{\Omega }}_2}} {\left( {x_1^{n - 4}x_2^{n - 2} + x_1^{n - 2}x_2^{n - 4} - 2x_1^{n - 3}x_2^{n - 3}} \right)d{x_1}d{x_2}}\\   
  &  = \int_0^{{{{\delta ^2}} \mathord{\left/
 {\vphantom {{{\delta ^2}} 2}} \right.
 \kern-\nulldelimiterspace} 2}} {\int_0^{{x_1}} {\left( {x_1^{n - 4}x_2^{n - 2} + x_1^{n - 2}x_2^{n - 4} - 2x_1^{n - 3}x_2^{n - 3}} \right)d{x_1}d{x_2}} }\\    
  &\quad {\text{   }} + \int_{{{{\delta ^2}} \mathord{\left/
 {\vphantom {{{\delta ^2}} 2}} \right.
 \kern-\nulldelimiterspace} 2}}^{{\delta ^2}} {\int_0^{\delta^2  - {x_1}} {\left( {x_1^{n - 4}x_2^{n - 2} + x_1^{n - 2}x_2^{n - 4} - 2x_1^{n - 3}x_2^{n - 3}} \right)d{x_1}d{x_2}} }\\    
  &  = \int_0^{{{{\delta ^2}} \mathord{\left/
 {\vphantom {{{\delta ^2}} 2}} \right.
 \kern-\nulldelimiterspace} 2}} {\frac{{x_1^{2n - 5}}}{{n - 1}} + \frac{{x_1^{2n - 5}}}{{n - 3}} - 2\frac{{x_1^{2n - 5}}}{{n - 2}}d{x_1}}\\   
  & \quad{\text{  }} + \int_{{{{\delta ^2}} \mathord{\left/
 {\vphantom {{{\delta ^2}} 2}} \right.
 \kern-\nulldelimiterspace} 2}}^{{\delta ^2}} {\frac{{x_1^{n - 4}{{\left( {{\delta ^2} - {x_1}} \right)}^{n - 1}}}}{{n - 1}} + \frac{{x_1^{n - 2}{{\left( {{\delta ^2} - {x_1}} \right)}^{n - 3}}}}{{n - 3}} - 2\frac{{x_1^{n - 3}{{\left( {{\delta ^2} - {x_1}} \right)}^{n - 2}}}}{{n - 2}}d{x_1}}\\   
  &  = \frac{{{{\left( {{{{\delta ^2}} \mathord{\left/
 {\vphantom {{{\delta ^2}} 2}} \right.
 \kern-\nulldelimiterspace} 2}} \right)}^{2n - 4}}}}{{\left( {n - 1} \right){{\left( {n - 2} \right)}^2}\left( {n - 3} \right)}} + K  \\
\end{split}
\end{equation}
\else
\begin{align}
\label{f66}
& \iint\limits_{{\bf{\Omega }}_1 + {{\bf{\Omega }}_2}} {x_1^{n - 4}x_2^{n - 4}{{\left( {{x_1} - {x_2}} \right)}^2}d{x_1}d{x_2}} \nonumber \\
& = \iint\limits_{{{\bf{\Omega }}_1} + {{\bf{\Omega }}_2}} {\left( {x_1^{n - 4}x_2^{n - 2} + x_1^{n - 2}x_2^{n - 4} - 2x_1^{n - 3}x_2^{n - 3}} \right)d{x_1}d{x_2}} \nonumber \\   
&  = \int_0^{{{{\delta ^2}} \mathord{\left/
 {\vphantom {{{\delta ^2}} 2}} \right.
 \kern-\nulldelimiterspace} 2}} \int_0^{{x_1}} ( x_1^{n - 4}x_2^{n - 2} + x_1^{n - 2}x_2^{n - 4} \nonumber \\ 
& \,\,\, - 2x_1^{n - 3}x_2^{n - 3} )d{x_1}d{x_2}  + \int_{{{{\delta ^2}} \mathord{\left/
 {\vphantom {{{\delta ^2}} 2}} \right.
 \kern-\nulldelimiterspace} 2}}^{{\delta ^2}} \int_0^{\delta^2  - {x_1}} ( x_1^{n - 4}x_2^{n - 2} \nonumber \\ 
& \,\,\, + x_1^{n - 2}x_2^{n - 4} - 2x_1^{n - 3}x_2^{n - 3} )d{x_1}d{x_2}  \nonumber \\    
&  = \int_0^{{{{\delta ^2}} \mathord{\left/
 {\vphantom {{{\delta ^2}} 2}} \right.
 \kern-\nulldelimiterspace} 2}} {\frac{{x_1^{2n - 5}}}{{n - 1}} + \frac{{x_1^{2n - 5}}}{{n - 3}} - 2\frac{{x_1^{2n - 5}}}{{n - 2}}d{x_1}} \nonumber \\   
& \,\,\, + \int_{{{{\delta ^2}} \mathord{\left/
 {\vphantom {{{\delta ^2}} 2}} \right.
 \kern-\nulldelimiterspace} 2}}^{{\delta ^2}} \frac{{x_1^{n - 4}{{\left( {{\delta ^2} - {x_1}} \right)}^{n - 1}}}}{{n - 1}} + \frac{{x_1^{n - 2}{{\left( {{\delta ^2} - {x_1}} \right)}^{n - 3}}}}{{n - 3}} \nonumber \\ 
& \,\,\, - 2\frac{{x_1^{n - 3}{{\left( {{\delta ^2} - {x_1}} \right)}^{n - 2}}}}{{n - 2}}d{x_1} \nonumber \\   
&  = \frac{{{{\left( {{{{\delta ^2}} \mathord{\left/
 {\vphantom {{{\delta ^2}} 2}} \right.
 \kern-\nulldelimiterspace} 2}} \right)}^{2n - 4}}}}{{\left( {n - 1} \right){{\left( {n - 2} \right)}^2}\left( {n - 3} \right)}} + K  \nonumber \\
\end{align}
\fi
where 
\ifCLASSOPTIONonecolumn
\begin{equation}
\label{f67}
K = \int_{{{{\delta ^2}} \mathord{\left/
 {\vphantom {{{\delta ^2}} 2}} \right.
 \kern-\nulldelimiterspace} 2}}^{{\delta ^2}} {\left( {\frac{{x_1^{n - 4}{{\left( {{\delta ^2} - {x_1}} \right)}^{n - 1}}}}{{n - 1}} + \frac{{x_1^{n - 2}{{\left( {\delta^2  - {x_1}} \right)}^{n - 3}}}}{{n - 3}} - 2\frac{{x_1^{n - 3}{{\left( {\delta^2  - {x_1}} \right)}^{n - 2}}}}{{n - 2}}} \right)d{x_1}}.
\end{equation}
\else
\begin{align}
\label{f67}
K & = \int_{{{{\delta ^2}} \mathord{\left/
 {\vphantom {{{\delta ^2}} 2}} \right.
 \kern-\nulldelimiterspace} 2}}^{{\delta ^2}} \left( \frac{{x_1^{n - 4}{{\left( {{\delta ^2} - {x_1}} \right)}^{n - 1}}}}{{n - 1}} + \frac{{x_1^{n - 2}{{\left( {\delta^2  - {x_1}} \right)}^{n - 3}}}}{{n - 3}} \right. \nonumber \\ & \left. \,\,\, - 2\frac{{x_1^{n - 3}{{\left( {\delta^2  - {x_1}} \right)}^{n - 2}}}}{{n - 2}} \right)d{x_1}.
\end{align}
\fi
\begin{figure}[!t]
\centering \includegraphics[width = 0.8\linewidth]{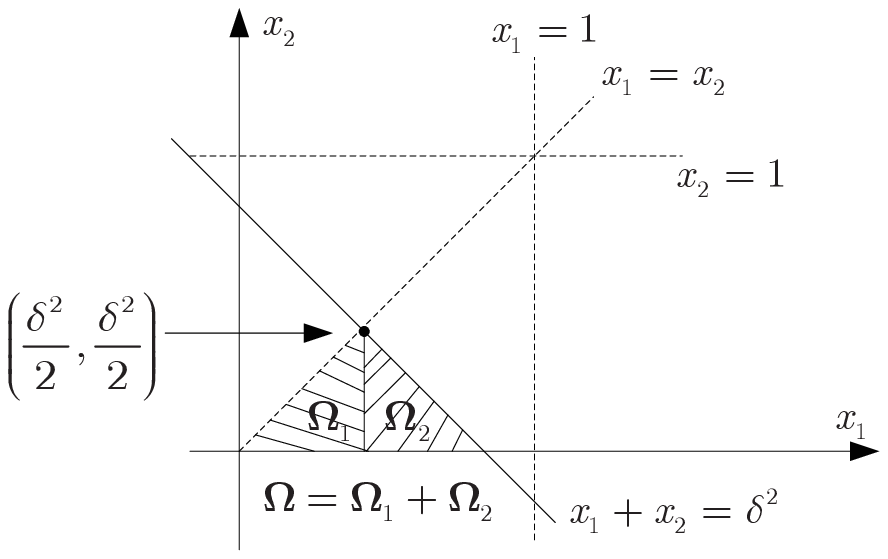}
\caption{{ The domain of integration in (\ref{f69})}}
\end{figure}

Let ${\delta ^2}y = {x_1}$, (\ref{f67}) could be further integrated as 
\ifCLASSOPTIONonecolumn
\begin{equation}
\label{f68}
\begin{split}
\quad \quad \quad \quad K = &\int_{{{{\delta ^2}} \mathord{\left/
 {\vphantom {{{\delta ^2}} 2}} \right.
 \kern-\nulldelimiterspace} 2}}^{{\delta ^2}} {\left( {\frac{{x_1^{n - 4}{{\left( {{\delta ^2} - {x_1}} \right)}^{n - 1}}}}{{n - 1}} + \frac{{x_1^{n - 2}{{\left( {\delta^2  - {x_1}} \right)}^{n - 3}}}}{{n - 3}} - 2\frac{{x_1^{n - 3}{{\left( {\delta^2  - {x_1}} \right)}^{n - 2}}}}{{n - 2}}} \right)d{x_1}}\\ 
& ={\delta ^{4n - 8}}\left( {\int_0^1 {\left( {\frac{{{y^{n - 4}}{{\left( {1 - y} \right)}^{n - 1}}}}{{n - 1}} + \frac{{{y^{n - 2}}{{\left( {1 - y} \right)}^{n - 3}}}}{{n - 3}} - 2\frac{{{y^{n - 3}}{{\left( {\delta  - y} \right)}^{n - 2}}}}{{n - 2}}} \right)dy} } \right)\\
  &\quad  - {\delta ^{4n - 8}}\left( {\int_0^{{1 \mathord{\left/
 {\vphantom {1 2}} \right.
 \kern-\nulldelimiterspace} 2}} {\left( {\frac{{{y^{n - 4}}{{\left( {1 - y} \right)}^{n - 1}}}}{{n - 1}} + \frac{{{y^{n - 2}}{{\left( {1 - y} \right)}^{n - 3}}}}{{n - 3}} - 2\frac{{{y^{n - 3}}{{\left( {\delta  - y} \right)}^{n - 2}}}}{{n - 2}}} \right)dy} } \right)\\
  &  = \frac{{{\delta ^{4n - 8}}}}{{n - 1}}\left( {B\left( {n - 3,n} \right) - B\left( {\frac{1}{2},n - 3,n} \right)} \right) \\
  &\quad + \frac{{{\delta ^{4n - 8}}}}{{n - 3}}\left( {B\left( {n - 1,n - 2} \right) - B\left( {\frac{1}{2},n - 1,n - 2} \right)} \right)\\
  &\quad  - \frac{{2{\delta ^{4n - 8}}}}{{n - 2}}\left( {B\left( {n - 2,n - 1} \right) - B\left( {\frac{1}{2},n - 2,n - 1} \right)} \right).
\end{split}
\end{equation}
\else
\begin{align}
\label{f68}
K & = \int_{{{{\delta ^2}} \mathord{\left/
 {\vphantom {{{\delta ^2}} 2}} \right.
 \kern-\nulldelimiterspace} 2}}^{{\delta ^2}} \left( \frac{{x_1^{n - 4}{{\left( {{\delta ^2} - {x_1}} \right)}^{n - 1}}}}{{n - 1}} + \frac{{x_1^{n - 2}{{\left( {\delta^2  - {x_1}} \right)}^{n - 3}}}}{{n - 3}} \right. \nonumber \\ 
 & \left. \,\,\, - 2\frac{{x_1^{n - 3}{{\left( {\delta^2  - {x_1}} \right)}^{n - 2}}}}{{n - 2}} \right)d{x_1} \nonumber \\ 
& ={\delta ^{4n - 8}}\left( \int_0^1 \left( \frac{{{y^{n - 4}}{{\left( {1 - y} \right)}^{n - 1}}}}{{n - 1}} + \frac{{{y^{n - 2}}{{\left( {1 - y} \right)}^{n - 3}}}}{{n - 3}} \right. \right. \nonumber \\ 
 & \left. \left. \,\,\, - 2\frac{{{y^{n - 3}}{{\left( {\delta  - y} \right)}^{n - 2}}}}{{n - 2}} \right)dy  \right) \nonumber \\
  & \,\,\, - {\delta ^{4n - 8}}\left( \int_0^{{1 \mathord{\left/
 {\vphantom {1 2}} \right. \kern-\nulldelimiterspace} 2}} \left( \frac{{{y^{n - 4}}{{\left( {1 - y} \right)}^{n - 1}}}}{{n - 1}} + \frac{{{y^{n - 2}}{{\left( {1 - y} \right)}^{n - 3}}}}{{n - 3}} \right. \right. \nonumber \\ 
 & \left. \left. \,\,\, - 2\frac{{{y^{n - 3}}{{\left( {\delta  - y} \right)}^{n - 2}}}}{{n - 2}} \right)dy  \right)\nonumber \\
  &  = \frac{{{\delta ^{4n - 8}}}}{{n - 1}}\left( {B\left( {n - 3,n} \right) - B\left( {\frac{1}{2},n - 3,n} \right)} \right) \nonumber \\
  &\quad + \frac{{{\delta ^{4n - 8}}}}{{n - 3}}\left( {B\left( {n - 1,n - 2} \right) - B\left( {\frac{1}{2},n - 1,n - 2} \right)} \right)\nonumber \\
  &\quad  - \frac{{2{\delta ^{4n - 8}}}}{{n - 2}}\left( {B\left( {n - 2,n - 1} \right) - B\left( {\frac{1}{2},n - 2,n - 1} \right)} \right).
\end{align}
\fi
When $1 < {\delta ^2} \leq 2$, the domain of integration is the set ${\bf{\Omega }} = \left\{ {{x_1} + {x_2} < \delta ,0 < {x_1},{x_2} < 1,1 < \delta  \leq 2} \right\}$ as shown in Fig. 18. Similar to the process of $0 \leq {\delta ^2} \leq 1$, the integration is written as 
\ifCLASSOPTIONonecolumn
\begin{equation}
\label{f69}
\begin{split}
&\iint\limits_{{{\bf{\Omega }}_1} + {{\bf{\Omega }}_2}} {x_1^{n - 4}x_2^{n - 4}{{\left( {{x_1} - {x_2}} \right)}^2}d{x_1}d{x_2}}\\
  &  = \iint\limits_{{{\bf{\Omega }}_1} + {{\bf{\Omega }}_2}} {\left( {x_1^{n - 4}x_2^{n - 2} + x_1^{n - 2}x_2^{n - 4} - 2x_1^{n - 3}x_2^{n - 3}} \right)d{x_1}d{x_2}}  \\ 
  &  = \int_0^{{{{\delta ^2}} \mathord{\left/
 {\vphantom {{{\delta ^2}} 2}} \right.
 \kern-\nulldelimiterspace} 2}} {\int_0^{{x_1}} {\left( {x_1^{n - 4}x_2^{n - 2} + x_1^{n - 2}x_2^{n - 4} - 2x_1^{n - 3}x_2^{n - 3}} \right)d{x_1}d{x_2}} }   \\ 
  & \quad {\text{   }} + \int_{{{{\delta ^2}} \mathord{\left/
 {\vphantom {{{\delta ^2}} 2}} \right.
 \kern-\nulldelimiterspace} 2}}^1 {\int_0^{\delta^2  - {x_1}} {\left( {x_1^{n - 4}x_2^{n - 2} + x_1^{n - 2}x_2^{n - 4} - 2x_1^{n - 3}x_2^{n - 3}} \right)d{x_1}d{x_2}} }.
\end{split}
\end{equation}
\else
\begin{align}
\label{f69}
& \iint\limits_{{{\bf{\Omega }}_1} + {{\bf{\Omega }}_2}} {x_1^{n - 4}x_2^{n - 4}{{\left( {{x_1} - {x_2}} \right)}^2}d{x_1}d{x_2}} \nonumber \\
  &  = \iint\limits_{{{\bf{\Omega }}_1} + {{\bf{\Omega }}_2}} {\left( {x_1^{n - 4}x_2^{n - 2} + x_1^{n - 2}x_2^{n - 4} - 2x_1^{n - 3}x_2^{n - 3}} \right)d{x_1}d{x_2}} \nonumber  \\ 
  &  = \int_0^{{{{\delta ^2}} \mathord{\left/
 {\vphantom {{{\delta ^2}} 2}} \right.
 \kern-\nulldelimiterspace} 2}} \int_0^{{x_1}} \left( x_1^{n - 4}x_2^{n - 2} + x_1^{n - 2}x_2^{n - 4} \right. \nonumber \\ 
 & \left. \,\,\, - 2x_1^{n - 3}x_2^{n - 3} \right)d{x_1}d{x_2} + \int_{{{{\delta ^2}} \mathord{\left/
 {\vphantom {{{\delta ^2}} 2}} \right.
 \kern-\nulldelimiterspace} 2}}^1 \int_0^{\delta^2  - {x_1}} \left( x_1^{n - 4}x_2^{n - 2} \right. \nonumber \\ 
 & \left. \,\,\, + x_1^{n - 2}x_2^{n - 4} - 2x_1^{n - 3}x_2^{n - 3} \right)d{x_1}d{x_2}.
\end{align}
\fi
The first term on the right hand of the second equation is the same as the case of $0 \leq \delta  \leq 1$, so it could be written as 
\ifCLASSOPTIONonecolumn
\begin{equation}
\label{f70}
\begin{split}
\iint\limits_{{{\bf{\Omega }}_1} + {{\bf{\Omega }}_2}} {x_1^{n - 4}x_2^{n - 4}{{\left( {{x_1} - {x_2}} \right)}^2}d{x_1}d{x_2}} = \frac{{{{\left( {{{{\delta ^2}} \mathord{\left/
 {\vphantom {{{\delta ^2}} 2}} \right.
 \kern-\nulldelimiterspace} 2}} \right)}^{2n - 4}}}}{{\left( {n - 1} \right){{\left( {n - 2} \right)}^2}\left( {n - 3} \right)}} + {K_1}
\end{split}
\end{equation}
\else
\begin{align}
\label{f70}
& \iint\limits_{{{\bf{\Omega }}_1} + {{\bf{\Omega }}_2}} {x_1^{n - 4}x_2^{n - 4}{{\left( {{x_1} - {x_2}} \right)}^2}d{x_1}d{x_2}} \nonumber \\ & = \frac{{{{\left( {{{{\delta ^2}} \mathord{\left/
 {\vphantom {{{\delta ^2}} 2}} \right.
 \kern-\nulldelimiterspace} 2}} \right)}^{2n - 4}}}}{{\left( {n - 1} \right){{\left( {n - 2} \right)}^2}\left( {n - 3} \right)}} + {K_1}
\end{align}
\fi
where
\ifCLASSOPTIONonecolumn
\begin{equation}
\label{f71}
\begin{split}
{K_1} &= \int_{{{{\delta ^2}} \mathord{\left/
 {\vphantom {{{\delta ^2}} 2}} \right.
 \kern-\nulldelimiterspace} 2}}^1 {\int_0^{\delta^2  - {x_1}} {\left( {x_1^{n - 4}x_2^{n - 2} + x_1^{n - 2}x_2^{n - 4} - 2x_1^{n - 3}x_2^{n - 3}} \right)d{x_1}d{x_2}} }   \\
  &  = \int_{{{{\delta ^2}} \mathord{\left/
 {\vphantom {{{\delta ^2}} 2}} \right.
 \kern-\nulldelimiterspace} 2}}^1 {\left( {\frac{{x_1^{n - 4}{{\left( {{\delta ^2} - {x_1}} \right)}^{n - 1}}}}{{n - 1}} + \frac{{x_1^{n - 2}{{\left( {\delta^2  - {x_1}} \right)}^{n - 3}}}}{{n - 3}} - 2\frac{{x_1^{n - 3}{{\left( {\delta^2  - {x_1}} \right)}^{n - 2}}}}{{n - 2}}} \right)d{x_1}}   \\
  &  = {\delta ^{4n - 8}}\int_0^1 {\left( {\frac{{{y^{n - 4}}{{\left( {1 - y} \right)}^{n - 1}}}}{{n - 1}} + \frac{{{y^{n - 2}}{{\left( {1 - y} \right)}^{n - 3}}}}{{n - 3}} - 2\frac{{{y^{n - 3}}{{\left( {1  - y} \right)}^{n - 2}}}}{{n - 2}}} \right)dy}   \\ 
  & \quad  - {\delta ^{4n - 8}}\int_0^{{1 \mathord{\left/
 {\vphantom {1 2}} \right.
 \kern-\nulldelimiterspace} 2}} {\left( {\frac{{{y^{n - 4}}{{\left( {1 - y} \right)}^{n - 1}}}}{{n - 1}} + \frac{{{y^{n - 2}}{{\left( {1 - y} \right)}^{n - 3}}}}{{n - 3}} - 2\frac{{{y^{n - 3}}{{\left( {1  - y} \right)}^{n - 2}}}}{{n - 2}}} \right)dy}   \\ 
  & \quad  - {\delta ^{4n - 8}}\int_0^1 {\left( {\frac{{{y^{n - 4}}{{\left( {1 - y} \right)}^{n - 1}}}}{{n - 1}} + \frac{{{y^{n - 2}}{{\left( {1 - y} \right)}^{n - 3}}}}{{n - 3}} - 2\frac{{{y^{n - 3}}{{\left( {1  - y} \right)}^{n - 2}}}}{{n - 2}}} \right)dy}   \\ 
  &\quad  + {\delta ^{4n - 8}}\int_0^{{1 \mathord{\left/
 {\vphantom {1 {{\delta ^2}}}} \right.
 \kern-\nulldelimiterspace} {{\delta ^2}}}} {\left( {\frac{{{y^{n - 4}}{{\left( {1 - y} \right)}^{n - 1}}}}{{n - 1}} + \frac{{{y^{n - 2}}{{\left( {1 - y} \right)}^{n - 3}}}}{{n - 3}} - 2\frac{{{y^{n - 3}}{{\left( {1  - y} \right)}^{n - 2}}}}{{n - 2}}} \right)dy}   \\ 
  &  = \frac{{{\delta ^{4n - 8}}}}{{n - 1}}\left( {B\left( {\frac{1}{{{\delta ^2}}},n - 3,n} \right) - B\left( {\frac{1}{2},n - 3,n} \right)} \right) \\
  & \quad + \frac{{{\delta ^{4n - 8}}}}{{n - 3}}\left( {B\left( {\frac{1}{{{\delta ^2}}},n - 1,n - 2} \right) - B\left( {\frac{1}{2},n - 1,n - 2} \right)} \right)  \\ 
  & \quad - \frac{{2{\delta ^{4n - 8}}}}{{n - 2}}\left( {B\left( {\frac{1}{{{\delta ^2}}},n - 2,n - 1} \right) - B\left( {\frac{1}{2},n - 2,n - 1} \right)} \right).
\end{split}
\end{equation}
\else
\begin{align}
\label{f71}
{K_1} &= \int_{{{{\delta ^2}} \mathord{\left/
 {\vphantom {{{\delta ^2}} 2}} \right.
 \kern-\nulldelimiterspace} 2}}^1 \int_0^{\delta^2  - {x_1}} \left( x_1^{n - 4}x_2^{n - 2} + x_1^{n - 2}x_2^{n - 4} \right. \nonumber \\ 
 & \left. \,\,\, - 2x_1^{n - 3}x_2^{n - 3} \right)d{x_1}d{x_2}  \nonumber \\
  &  = \int_{{{{\delta ^2}} \mathord{\left/
 {\vphantom {{{\delta ^2}} 2}} \right.
 \kern-\nulldelimiterspace} 2}}^1 \left( \frac{{x_1^{n - 4}{{\left( {{\delta ^2} - {x_1}} \right)}^{n - 1}}}}{{n - 1}} + \frac{{x_1^{n - 2}{{\left( {\delta^2  - {x_1}} \right)}^{n - 3}}}}{{n - 3}} \right. \nonumber \\ 
 & \left. \,\,\, - 2\frac{{x_1^{n - 3}{{\left( {\delta^2  - {x_1}} \right)}^{n - 2}}}}{{n - 2}} \right)d{x_1} \nonumber \\ 
 & = {\delta ^{4n - 8}}\int_0^1 \left( \frac{{{y^{n - 4}}{{\left( {1 - y} \right)}^{n - 1}}}}{{n - 1}} + \frac{{{y^{n - 2}}{{\left( {1 - y} \right)}^{n - 3}}}}{{n - 3}} \right. \nonumber \\ 
 & \left. \,\,\, - 2\frac{{{y^{n - 3}}{{\left( {1  - y} \right)}^{n - 2}}}}{{n - 2}} \right)dy \nonumber \\ 
 & \,\,\, - {\delta ^{4n - 8}}\int_0^{{1 \mathord{\left/
 {\vphantom {1 2}} \right.
 \kern-\nulldelimiterspace} 2}} \left( \frac{{{y^{n - 4}}{{\left( {1 - y} \right)}^{n - 1}}}}{{n - 1}} + \frac{{{y^{n - 2}}{{\left( {1 - y} \right)}^{n - 3}}}}{{n - 3}} \right. \nonumber \\ 
 & \left. \,\,\,- 2\frac{{{y^{n - 3}}{{\left( {1  - y} \right)}^{n - 2}}}}{{n - 2}} \right)dy   \\ 
  &\,\,\,  - {\delta ^{4n - 8}}\int_0^1 \left( \frac{{{y^{n - 4}}{{\left( {1 - y} \right)}^{n - 1}}}}{{n - 1}} + \frac{{{y^{n - 2}}{{\left( {1 - y} \right)}^{n - 3}}}}{{n - 3}} \right. \nonumber \\ 
 & \left. \,\,\, - 2\frac{{{y^{n - 3}}{{\left( {1  - y} \right)}^{n - 2}}}}{{n - 2}} \right)dy  \nonumber \\ 
  &\,\,\, + {\delta ^{4n - 8}}\int_0^{{1 \mathord{\left/
 {\vphantom {1 {{\delta ^2}}}} \right.
 \kern-\nulldelimiterspace} {{\delta ^2}}}} \left( \frac{{{y^{n - 4}}{{\left( {1 - y} \right)}^{n - 1}}}}{{n - 1}} + \frac{{{y^{n - 2}}{{\left( {1 - y} \right)}^{n - 3}}}}{{n - 3}} \right. \nonumber \\ 
 & \left. \,\,\, - 2\frac{{{y^{n - 3}}{{\left( {1  - y} \right)}^{n - 2}}}}{{n - 2}} \right)dy \nonumber \\ 
  & = \frac{{{\delta ^{4n - 8}}}}{{n - 1}}\left( {B\left( {\frac{1}{{{\delta ^2}}},n - 3,n} \right) - B\left( {\frac{1}{2},n - 3,n} \right)} \right) \nonumber \\
  & \,\,\, + \frac{{{\delta ^{4n - 8}}}}{{n - 3}}\left( B\left( {\frac{1}{{{\delta ^2}}},n - 1,n - 2} \right) \right. \nonumber \\ 
 & \left. \,\,\, - B\left( {\frac{1}{2},n - 1,n - 2} \right) \right) \nonumber \\ 
 & \,\,\, - \frac{{2{\delta ^{4n - 8}}}}{{n - 2}}\left( B\left( {\frac{1}{{{\delta ^2}}},n - 2,n - 1} \right) \right. \nonumber \\ 
 & \left. \,\,\, - B\left( {\frac{1}{2},n - 2,n - 1} \right) \right).
\end{align}
\fi
Let $\bar B\left( {\alpha ,m,n} \right) = B\left( {m,n} \right) - B\left( {\alpha ,m,n} \right)$ and ${\bar B_1}\left( {\alpha ,m,n} \right) = \bar B\left( {\alpha ,m,n} \right) - \bar B\left( {\frac{1}{2},m,n} \right)$ , combining the constant $M$, (\ref{f66}), and (\ref{f70}), (65) could be written as the final formula
\begin{equation}
\label{f72}
p\left( {{d_{pF}} \leq \delta } \right) = \left\{ {\begin{array}{*{20}{c}}
  {{{\left( {\frac{\delta }{2}} \right)}^{2n - 4}} + M{\delta ^{4n - 8}}{P_1}\left( n \right)} \\ 
  {{{\left( {\frac{\delta }{2}} \right)}^{2n - 4}} + M{\delta ^{4n - 8}}{P_2}\left( n \right)} 
\end{array}} \right.
\end{equation}
where
\ifCLASSOPTIONonecolumn
$$
{P_1}\left( n \right) = \frac{1}{{n - 3}}\bar B\left( {\frac{1}{2},n - 1,n - 2} \right) + \frac{1}{{n - 1}}\bar B\left( {\frac{1}{2},n - 3,n} \right) - \frac{2}{{n - 2}}\bar B\left( {\frac{1}{2},n - 2,n - 1} \right)
$$
\else
\begin{align*}
{P_1}\left( n \right) & = \frac{1}{{n - 3}}\bar B\left( {\frac{1}{2},n - 1,n - 2} \right) \\ 
& \,\,\, + \frac{1}{{n - 1}}\bar B\left( {\frac{1}{2},n - 3,n} \right) \\ 
& \,\,\, - \frac{2}{{n - 2}}\bar B\left( {\frac{1}{2},n - 2,n - 1} \right)
\end{align*}
\fi
and
\ifCLASSOPTIONonecolumn
$$
 {P_2}\left( n \right) = \frac{1}{{n - 3}}{{\bar B}_1}\left( {\frac{1}{{{\delta ^2}}},n - 1,n - 2} \right) + \frac{1}{{n - 1}}{{\bar B}_1}\left( {\frac{1}{{{\delta ^2}}},n - 3,n} \right) - \frac{2}{{n - 2}}{{\bar B}_1}\left( {\frac{1}{{{\delta ^2}}},n - 2,n - 1} \right)
$$
\else
\begin{align*}
 {P_2}\left( n \right) & = \frac{1}{{n - 3}}{{\bar B}_1}\left( {\frac{1}{{{\delta ^2}}},n - 1,n - 2} \right) \\ 
& \,\,\, + \frac{1}{{n - 1}}{{\bar B}_1}\left( {\frac{1}{{{\delta ^2}}},n - 3,n} \right) \\ 
& \,\,\, - \frac{2}{{n - 2}}{{\bar B}_1}\left( {\frac{1}{{{\delta ^2}}},n - 2,n - 1} \right)
\end{align*}
\fi
\end{IEEEproof}





\ifCLASSOPTIONcaptionsoff
  \newpage
\fi

\end{document}